\documentclass[longauth]{aa}
\usepackage{gensymb}
\usepackage{graphicx}
\usepackage{txfonts}
\usepackage{xcolor}
\usepackage{hyperref}
\begin{document}

   \title{The NIKA2 cosmological legacy survey at 2\,mm: catalogs, colors, redshift distributions, and implications for deep surveys}
   
    \authorrunning{B\'ethermin et al.}
    \titlerunning{N2CLS 2\,mm sources}

   \author{M.~B\'ethermin \inst{\ref{Strasbourg}} 
     \and  G.~Lagache \inst{\ref{LAM}}
     \and  C.~Carvajal-Bohorquez \inst{\ref{LAM}}
     \and R.~Adam \inst{\ref{OCA}}
     \and  P.~Ade \inst{\ref{Cardiff}}
     \and  H.~Ajeddig \inst{\ref{CEA}}
     \and  S.~Amarantidis \inst{\ref{IRAME}}
     \and  P.~Andr\'e \inst{\ref{CEA}}
     \and  H.~Aussel \inst{\ref{CEA}}
     \and  A.~Beelen \inst{\ref{LAM}}
     \and  A.~Beno\^it \inst{\ref{Neel}}
     \and  S.~Berta \inst{\ref{IRAMF}}
     \and  L.~J.~Bing \inst{\ref{Sussex}}
     \and  A.~Bongiovanni \inst{\ref{IRAME}}
     \and  J.~Bounmy \inst{\ref{LPSC}}
     \and  O.~Bourrion \inst{\ref{LPSC}}
     \and  M.~Calvo \inst{\ref{Neel}}
     \and  A.~Catalano \inst{\ref{LPSC}}
     \and  D.~Ch\'erouvrier \inst{\ref{LPSC}}
     \and  M.~De~Petris \inst{\ref{Roma}}
     \and  F.-X.~D\'esert \inst{\ref{IPAG}}
     \and  S.~Doyle \inst{\ref{Cardiff}}
     \and  E.~F.~C.~Driessen \inst{\ref{IRAMF}}
     \and  G.~Ejlali \inst{\ref{Teheran}}
     \and  A.~Ferragamo \inst{\ref{Roma}}
     \and  A.~Gomez \inst{\ref{CAB}} 
     \and  J.~Goupy \inst{\ref{Neel}}
     \and  C.~Hanser \inst{\ref{CPPM}}
     \and  S.~Katsioli \inst{\ref{AthenObs}, \ref{AthenUniv}}
     \and  F.~K\'eruzor\'e \inst{\ref{Argonne}}
     \and  C.~Kramer \inst{\ref{IRAMF}}
     \and  B.~Ladjelate \inst{\ref{IRAME}} 
     \and  S.~Leclercq \inst{\ref{IRAMF}}
     \and  J.-F.~Lestrade \inst{\ref{LERMA}}
     \and  J.~F.~Mac\'ias-P\'erez \inst{\ref{LPSC}}
     \and  S.~C.~Madden \inst{\ref{CEA}}
     \and  A.~Maury \inst{\ref{Barcelona1}, \ref{Barcelona2}, \ref{CEA}}
     \and  F.~Mayet \inst{\ref{LPSC}}
     \and  A.~Monfardini \inst{\ref{Neel}}
     \and  A.~Moyer-Anin \inst{\ref{LPSC}}
     \and  M.~Mu\~noz-Echeverr\'ia \inst{\ref{IRAP}}
     \and  I.~Myserlis \inst{\ref{IRAME}}
     \and  A.~Paliwal \inst{\ref{Roma2}}
     \and  L.~Perotto \inst{\ref{LPSC}}
     \and  G.~Pisano \inst{\ref{Roma}}
     \and  N.~Ponthieu \inst{\ref{IPAG}}
     \and  V.~Rev\'eret \inst{\ref{CEA}}
     \and  A.~J.~Rigby \inst{\ref{Leeds}}
     \and  A.~Ritacco \inst{\ref{LPSC}}
     \and  H.~Roussel \inst{\ref{IAP}}
     \and  F.~Ruppin \inst{\ref{IP2I}}
     \and  M.~S\'anchez-Portal \inst{\ref{IRAME}}
     \and  S.~Savorgnano \inst{\ref{LPSC}}
     \and  K.~Schuster \inst{\ref{IRAMF}}
     \and  A.~Sievers \inst{\ref{IRAME}}
     \and  C.~Tucker \inst{\ref{Cardiff}}
     \and  R.~Zylka \inst{\ref{IRAMF}}
     }
   \institute{Universit\'e de Strasbourg, CNRS, Observatoire astronomique de Strasbourg, UMR 7550, 67000 Strasbourg, France, \email{matthieu.bethermin@astro.unistra.fr} 
     \label{Strasbourg}
    \and	
     Aix Marseille Univ, CNRS, CNES, LAM (Laboratoire d'Astrophysique de Marseille), Marseille, France
     \label{LAM}
     \and
     Universit\'e C\^ote d'Azur, Observatoire de la C\^ote d'Azur, CNRS, Laboratoire Lagrange, France 
     \label{OCA}
     \and
     School of Physics and Astronomy, Cardiff University, Queen’s Buildings, The Parade, Cardiff, CF24 3AA, UK 
     \label{Cardiff}
     \and
     Université Paris Cité, Université Paris-Saclay, CEA, CNRS, AIM, F-91191 Gif-sur-Yvette, France
     \label{CEA}
     \and
     Institut de Radioastronomie Millim\'etrique (IRAM), Avenida Divina Pastora 7, Local 20, E-18012, Granada, Spain
     \label{IRAME}     
     \and
     Institut N\'eel, CNRS, Universit\'e Grenoble Alpes, France
     \label{Neel}
     \and
     Astronomy Centre, Department of Physics and Astronomy, University of Sussex, Brighton BN1 9QH
     \label{Sussex}
     \and 
     Institut de RadioAstronomie Millim\'etrique (IRAM), Grenoble, France
     \label{IRAMF}
     \and
      Univ. Grenoble Alpes, CNRS, Grenoble INP, LPSC-IN2P3, 53, avenue des Martyrs, 38000 Grenoble, France
     \label{LPSC}
     \and 
     Dipartimento di Fisica, Sapienza Universit\`a di Roma, Piazzale Aldo Moro 5, I-00185 Roma, Italy
     \label{Roma}
     \and
     Univ. Grenoble Alpes, CNRS, IPAG, 38000 Grenoble, France 
     \label{IPAG}
     \and
     Institute for Research in Fundamental Sciences (IPM), School of Astronomy, Tehran, Iran
     \label{Teheran}
     \and
     Centro de Astrobiolog\'ia (CSIC-INTA), Torrej\'on de Ardoz, 28850 Madrid, Spain
     \label{CAB}
     \and
     Aix Marseille Univ, CNRS/IN2P3, CPPM, Marseille, France
     \label{CPPM}
     \and
     National Observatory of Athens, Institute for Astronomy, Astrophysics, Space Applications and Remote Sensing, Ioannou Metaxa
     and Vasileos Pavlou GR-15236, Athens, Greece
     \label{AthenObs}
     \and
     Department of Astrophysics, Astronomy \& Mechanics, Faculty of Physics, University of Athens, Panepistimiopolis, GR-15784
     Zografos, Athens, Greece
     \label{AthenUniv}
     \and
     High Energy Physics Division, Argonne National Laboratory, 9700 South Cass Avenue, Lemont, IL 60439, USA
     \label{Argonne}
     \and  
     LERMA, Observatoire de Paris, PSL Research University, CNRS, Sorbonne Universit\'e, UPMC, 75014 Paris, France  
     \label{LERMA}
     \and
     Institute of Space Sciences (ICE), CSIC, Campus UAB, Carrer de Can Magrans s/n, E-08193, Barcelona, Spain
     \label{Barcelona1}
     \and
     ICREA, Pg. Lluís Companys 23, Barcelona, Spain
     \label{Barcelona2}
     \and
     IRAP, CNRS, Université de Toulouse, CNES, UT3-UPS, (Toulouse), France 
     \label{IRAP}
     \and
     Dipartimento di Fisica, Universit\`a di Roma ‘Tor Vergata’, Via della Ricerca Scientifica 1, I-00133 Roma, Italy	
     \label{Roma2}
     \and
     School of Physics and Astronomy, University of Leeds, Leeds LS2 9JT, UK
     \label{Leeds}
     \and
     Laboratoire de Physique de l’\'Ecole Normale Sup\'erieure, ENS, PSL Research University, CNRS, Sorbonne Universit\'e, Universit\'e de Paris, 75005 Paris, France 
     \label{ENS}
     \and
     INAF-Osservatorio Astronomico di Cagliari, Via della Scienza 5, 09047 Selargius, IT
     \label{INAF}
     \and    
     Institut d'Astrophysique de Paris, CNRS (UMR7095), 98 bis boulevard Arago, 75014 Paris, France
     \label{IAP}
     \and
     University of Lyon, UCB Lyon 1, CNRS/IN2P3, IP2I, 69622 Villeurbanne, France
     \label{IP2I}
     \and
     University Federico II, Naples, Italy
     \label{Naples}
   }
     
    \date{Received XXX; accepted XXX}

 
  \abstract{}
  {Millimeter galaxy surveys are particularly effective in detecting dusty star-forming galaxies at high redshift. While such observations are typically conducted at $\sim$1\,mm, various authors have suggested over the last 20 years that 2\,mm observations may be better suited for selecting sources at even higher redshifts. In this paper, we use the unprecedented 2\,mm data from the NIKA2 Cosmological Legacy Survey (N2CLS), together with the simulated infrared dusty extragalactic sky (SIDES), to study and interpret the statistical properties of galaxies selected at this wavelength.}
  {We use the N2CLS robust sample (95\,\% purity) at 2\,mm ($\sim$18\,arcsec resolution), which contains 25 sources in the deep GOODS-N field (159\,arcmin$^2$, 0.047\,mJy/beam RMS) and 90 sources in the wide COSMOS field (1130\,arcmin$^2$, 0.09\,mJy/beam RMS). The sources are matched with the N2CLS 1.2\,mm sources, the ancillary 850\,$\mu$m sources, and redshift catalogs to study the colors and redshift distributions. We also produce end-to-end simulations based on SIDES and the observed N2CLS detector timelines to interpret the data.}
  {We find a mean S$_{\rm 2\,mm}$/S$_{\rm 1.2\,mm}$ color of 0.215$\pm$0.006 with a standard deviation of 0.056$\pm$0.004, which is mainly caused by resolution and source extraction effects according to the SIDES simulation. We measure a mean redshift of 3.6$\pm$0.3 in GOODS-N, which is marginally higher than expectations from SIDES (3.0$\pm$0.2) because of an overdensity at z$\sim$5.2, and 3.2$\pm$0.2 in COSMOS, which agrees with the 3.2$\pm$0.2 predicted by SIDES. We also show that the observed S$_{\rm 2\,mm}$/S$_{\rm 1.2\,mm}$ colors exhibit a weak dependence with redshift but a large dispersion, which limits its efficiency to select high-z sources. We compare the measured fluxes of 2\,mm sources detected by both the N2CLS and the Ex-MORA surveys. The results only agree if we take into account the impact of the bandwidth, the source blending, and the source size. Finally, we study the eight 2\,mm sources not detected at 1.2\,mm, and found that two of them are radiogalaxies, one is a $z\sim2$ galaxy, and the remaining 6 (all in COSMOS) are compatible with the expected number of spurious detections. Consequently, the N2CLS survey shows no evidence for any exotic dusty galaxy population with either very cold dust or at very high redshift, which would be detected only at 2\,mm.  Using SIDES, we show that 2\,mm samples have a higher mean redshift compared to 1.2\,mm because they miss dusty galaxies around cosmic noon (z$\sim$2). Finally, we discuss the efficiency of single-dish and interferometric blind surveys to build samples of high-z dusty galaxies. The data can be accessed at \url{https://data.lam.fr/n2cls/}.}
   {}
%

\maketitle

\nolinenumbers


\section{Introduction}

Since the discovery of high-redshift bright submillimeter galaxies \citep[e.g,][]{Smail1997,Hughes1998,Blain2002,Chapman2003}, the (sub)milimeter atmospheric window is the prime wavelength range to probe dusty star-forming systems in the high-z Universe (z$>$2, see, e.g., \citealt{Casey2014} and \citealt{Hodge2020} for reviews). Indeed, the peak of cold dust emission in the far-infrared ($\sim$80\,$\mu$m) tracing obscured star formation is shifted to the sub-millimeter windows by the redshift. The observed submillimeter flux from galaxies at higher redshifts comes from rest-frame wavelengths closer from their peak of emission, compensating the increasing luminosity distance and leading to an almost constant flux-versus-luminosity ratio at z$\gtrsim$2 \citep[e.g.,][]{Blain2002,Lagache2004}. This effect is commonly referred to as the negative K-correction.

Observations unveiled a numerous population of these submillimeter galaxies challenging galaxy evolution models, which struggle to produce enough objects with such high star-formation rates in the early Universe \citep[e.g,][]{Baugh2005,Lovell2021,Hayward2021,Kumar2025}. The Atacama large millimeter array (ALMA) showed that massive dusty galaxies already exist up to z$\sim$7 \citep[e.g.,][]{Strandet2017,Fudamoto2021}. Several surveys have found that the obscured star formation density is already of the same order of magnitude as the unobscured one at z$\sim$5 \citep[e.g.,][]{Gruppioni2020,Zavala2021,Khusanova2021,Barrufet2023}. Because of their large dust content, which absorbs a large fraction of the light produced by their stars and re-emits it in the rest-frame far infrared, dusty galaxies are very faint in the optical \citep[e.g.,][]{Wang2019,Talia2021}, but very bright in the observed (sub)millimeter.

While the first surveys were performed at 850\,$\mu$m by the submillimeter common-user bolometer array (SCUBA, \citealt{Holland1999}), other instruments on single-dish telescopes were developed to probe longer wavelengths\footnote{This paper focuses on wavelengths that are strictly longer than 1\,mm.} such as Bolocam \citep{Laurent2005}, AzTEC \citep[][for a synthesis]{Scott2012}, or MAMBO \citep{Lindner2011} at 1.1/1.2\,mm, and GISMO \citep{Staguhn2014,Magnelli2019} at 2\,mm. In addition, a population of very bright sources at 1.4\,mm magnified by gravitational lensing was found by the south pole telescope (SPT, \citealt{Vieira2013}, see also \citealt{Lestrade2010} about an earlier single detection with MAMBO). Finally, ALMA offers capabilities to perform interferometric surveys through large mosaics of pointings, and blind photometric surveys were performed at 1\,mm \citep[e.g.,][]{Dunlop2017,Franco2018} and 2\,mm \citep[e.g.,][]{Casey2021,Long2024}.  The survey at longer wavelengths found sources at a higher average redshift \citep[e.g.,][]{Simpson2014,Weiss2013,Staguhn2014,Casey2021,Cowie2023}. This effect can easily be modeled considering the combination of the different  (sub)millimeter flux versus redshift relation at different wavelengths \citep{Zavala2014}, the strong evolution of the dust luminosity function, and the impact of lensing at high flux \citep{Bethermin2015}.
 
Performing deep surveys at longer wavelengths may appear an effective strategy for maximizing the number of high-redshift detections. However, things are more complex than they seem. At 2\,mm, dusty galaxies even up to z$\sim7$ are observed in the Rayleigh-Jeans regime with a flux density decreasing steeply with increasing wavelength. If the survey sensitivity is not significantly better than at shorter wavelength, this could lead to fewer detections even at high redshift. However, surveys at 2\,mm could be potentially powerful to detect hypothetical rare dusty objects at very high redshifts. They can also be combined with shorter wavelength surveys to identify follow-up targets at high redshift \citep[e.g.,][]{Casey2019,Cooper2022,Cowie2023} or with a very cold apparent dust temperature \citep[e.g,][]{Jin2022,Bing2024}.

The new IRAM KIDS array (NIKA2, \citealt{Monfardini2014,Calvo2016,Bourrion2016}) was installed on the IRAM 30-meter telescope in October 2015 \citep{Adam2018,Perotto2020} and can observe its circular 6.5\,arcmin-diameter field of view simultaneously at 1.2 and 2\,mm. Thanks to its combined high sensitivity and large field of view, NIKA2 is particularly efficient to perform blind surveys at these two wavelengths. The NIKA2 cosmological legacy survey (N2CLS) observed the GOODS-N and COSMOS fields. It provides both an unprecedented statistical sample of 2\,mm sources and an unique opportunity to combine them with their 1.2\,mm counterparts. First results based on data obtained before May 2021 are shown in \citet{Bing2023}, which extensively discusses the number counts. At that time, the GOODS-N survey was complete and COSMOS was still missing a factor 2 in integration time. Since then, the depth of COSMOS has increased, but this flux regime was already probed by GOODS-N. Therefore, the new data does not significantly improve the constraints on the number counts. For the GOODS-N field, the detailed analysis of individual sources is presented in \citet{Berta2025} and \citet{Lagache2025}, and the confusion due to undetected sources in \citet{Ponthieu2025}. A similar study is also performed on the serendipitous point-source detections in the NIKA2 large program targeting the Sunyaev-Zeldovich effect in galaxy clusters (LPSZ, D\'esert et al. in prep.).

In this paper, we present the final N2CLS 2\,mm data, interpret their color and redshift distributions, and discuss the consequences for 2\,mm surveys. In Sect.\,\ref{sect:data}, we introduce the observed (N2CLS) and simulated data used in the paper. We describe the 2\,mm catalog together with the methods used to produce it. In Sect.\,\ref{sect:properties}, we discuss the properties of the sources, such as their color and their redshift. We then discuss the relevance of 2\,mm single-dish and interferometric photometric surveys in identifying the high-redshift populations of dusty star-forming galaxies in Sect.\,\ref{sect:discussion}, and finally conclude in Sect.\,\ref{sect:conclusion}. 


\section{Data}

\label{sect:data}

\subsection{The NIKA2 cosmological legacy survey}

The N2CLS is a 300\,h guaranteed-time large program of the NIKA2 camera at the 30\,m IRAM telescope (192-16, PIs: Lagache, Beelen, and Ponthieu). Thanks to its unprecedented mapping speed for a single-dish millimeter instrument, NIKA2 allowed us to map two popular fields with unprecedented depth compared to other millimeter surveys of the same size in GOODS-N, and a similar depth but a twice larger area than the Ex-MORA survey in COSMOS \citep{Long2024}. For both fields, the data reduction and map making were performed using the pointing and imaging in continuum (PIIC) data reduction pipeline developed and supported by IRAM \citep{Zylka2013,piic2024}. The beam FWHM is $\sim$12\,arcsec and $\sim$18\,arcsec at 1.2\,mm and 2\,mm, respectively.

The GOODS-N field covers 159\,arcmin$^2$ with a 1\,$\sigma$ depth of 0.17\,mJy/beam and 0.047\,mJy/beam at 1.2\,mm and 2\,mm, respectively. All the data were taken between October 2017 and May 2021. After scan selection (i.e. removing scans with higher noise linked to weather instabilities), the total time on source reached 78.0\,h and 72.8\,h at 1.2 and 2\,mm respectively. The data reduction is described in \citet{Bing2023}. These data are close to the confusion limit \citep{Ponthieu2025}.

In the COSMOS field, we achieve a 1$\sigma$ sensitivity of 0.3\,mJy/beam at 1.2\,mm and 0.09\,mJy/beam at 2\,mm, covering an area of 1130\,arcmin$^2$. 
The data were taken between October 2017 and January 2023. After scan selection, the total observing time reached 195\,h. \citet{Bing2023} presented the analysis of the data obtained before May 2021 (representing on-source time of 78.7\,h at 1.2\,mm and and 79\,h at 2\,mm). We used the same method but with an updated version of PIIC to reduce the full dataset (Carvajal-Bohorquez et al. in prep. for the details).

The absolute calibration uncertainty, coming from the planet models used to calibrate the NIKA2 data, is 5\%. The point-source root-mean-square (RMS) calibration uncertainties are about 6\% at 1.2\,mm and about 3\% at 2\,mm \citep{Perotto2020}.

\subsection{The GOODS-N catalog}

\label{sect:goodsn_cat}

We use the 1.2\,mm and 2\,mm catalogs produced from NIKA2 and described in \citet{Bing2023}. The sources are extracted from match-filtered\footnote{Original maps are cross-correlated by the beam to maximize the S/N ratio of point sources and optimize their detection.} S/N maps applying a detection threshold. The source fluxes are then measured performing PSF-fitting photometry. End-to-end simulations (see Sect.\,\ref{sect:e2e}) provided estimates of the purity (fraction of real sources in the extracted sample) as a function of this threshold. \citet{Bing2023} used a S/N threshold corresponding to 80\,\% purity (3.0 at 1.2\,mm and 2.9 at 2\,mm). The biases impacting the photometry, such as flux boosting and filtering effects, are corrected using the median ratio between the intrinsic flux in the simulation and the output flux measured using the simulated map. The simulation are also used to compute the flux uncertainties, which are significantly larger than what we could expect from pure uncorrelated noise.

In this paper focused on 2\,mm sources, we select only high-reliability 2\,mm sources corresponding to a sample purity of 95\,\% (S/N$\ge$4.2). We obtain a sub-sample of 25 sources\footnote{\citet{Berta2025} reported 26 sources, but N2GN\_1\_26 has a S/N=4.18 in the latest version of the catalog, and is thus below our selection threshold.} that we then match with the full 1.2\,mm catalog (S/N$\ge$3.0 corresponding to 80\,\% purity) using a matching radius of 6.5\,arcsec. Only one object has no 1.2\,mm counterpart (N2GN\_2\_13)\footnote{The name of N2GN\_2\_20 could suggest that it is also a source detected only at 2\,mm, but it has a weak detection at 1.2\,mm catalog (SNR$_{\rm 1.2\,mm}=3.3$) complementing the robust 2\,mm detection (SNR$_{\rm 2mm}=5.1$).}. This source is close to the noisy edge of the field. We estimated a 5\,$\sigma$ upper limit on its flux based on the flux uncertainties of sources located in areas with a similar depth. The catalog is given in Table\,\ref{tab:cat_goodsn}.

The counterparts of all these sources were identified using a large set of ancillary data and dedicated NOEMA follow up observations. This counterpart identification and the redshift determination is described in a companion paper \citep{Berta2025}. This procedure also shows that 4 of our 25 sources are multiple (N2GN\_1\_12, N2GN\_1\_17, N2GN\_1\_24, N2GN\_1\_34).

\subsection{The COSMOS catalog}

\label{sect:cosmos_cat}

The observing time in the COSMOS field has been more than doubled since the source extraction of \citet{Bing2023}. We thus produced new 1.2\,mm and 2\,mm catalogs using the full-depth maps, together with updated end-to-end simulations to estimate the sample purity and correct for the flux boosting and filtering effects. The source extraction and flux measurement were conducted following the same approach as in GOODS-N. To estimate the purity, we compared the number of sources between end-to-end simulations and pure-noise maps to derive the fraction of spurious sources at a given S/N (see Carvajal-Bohorquez et al. for a full description of the method). We corrected the raw flux densities and estimated uncertainties using the same approach as in GOODS-N, with only minor changes. We estimate that a purity of 95\,\% is reached for a S/N threshold of 4.6, which we then use to produce our secure 2\,mm catalog containing 90\,objects. The details will be provided in Carvajal-Bohorquez et al. (in prep.). 

We use a similar matching procedure as in the GOODS-N field to associate the 2\,mm source with their 1.2\,mm counterparts. The 1.2\,mm catalog used for this procedure is cut at S/N$\ge$3.9, corresponding to 80\,\% purity. We find a 1.2\,mm counterpart for 82 of our 90 sources. Two of the 2\,mm sources (N2CO\_1\_8 and N2CO\_1\_41) split into two 1.2\,mm counterparts each thanks to the sharper beam at shorter wavelength. Two other sources (N2CO\_2\_29 and N2CO\_2\_61) are complex blends with a clear 1.2\,mm emission near the 2\,mm source. However, it is unclear what fraction of the 1.2\,mm emission comes from the main 2\,mm source. For instance, N2CO\_2\_61 is detected at 1.2\,mm, but it is an intriguing case of a complex blend between the faint 1.2\,mm emission associated with the radio source COSMOSVLADP\_J095951.93+020542.6 and a brighter 1.2\,mm dusty neighbor 8.8\,arcsec away (Carvajal-Bohorquez et al., in prep.). The final catalog is provided in Table\,\ref{tab:cat_cosmos} and \ref{tab:cat_cosmos2}.


To identify the multi-wavelength counterpart(s) corresponding to each N2CLS source, we first searched 
for a counterpart in ancillary high-resolution (sub-)millimeter interferometric catalogs, which serve as a proxy for accurate source position (following \citealt{Berta2025}). We used the following datasets, in decreasing order of priority:

\begin{itemize}
\item AS2COSMOS \citep{AS2COSMOS}, which  was  given highest  priority because of dedicated follow-ups with ALMA of the bright 850\,$\mu$m SMGs; 
\item A$^3$COSMOS \citep[][]{A3COSMOS}, a (sub-)millimeter ALMA archival survey comprising a prior catalog (S/N$\ge4.35$), and a complementary blind catalog (corresponding to sources detected blindly with SNR$\ge$5.4 and not present in the prior catalog). We used the latest available version of the catalogs\footnote{\url{https://sites.google.com/view/a3cosmos}} (2025/03/12).
\item the radio JVLA 3\,GHz catalog from \citet{3GHZVLA} if no counterpart is found from ALMA.
\end{itemize}

We visually inspected all our sources to verify the proxy positions and to check for any missing counterparts. During this process, we identified 3 MeerKAT sources \citep[MIGHTEE -- MeerKAT International Gigahertz Tiered Extragalactic Exploration survey;][]{Meerkat}.
After pinpointing the precise location of the N2CLS sources using these high-resolution data, we searched for their short-wavelength counterparts to determine a spectroscopic redshift when available, or a photometric redshift otherwise.
We obtained 8 spectroscopic redshifts from \citet{Chen2022}, complemented by \citet{Mitsuhashi_22_s2cosmos_zspec} from the brightest AS2COSMOS sources. We also collected 33 spectroscopic redshifts from the COSMOS spectroscopic redshift compilation by \citet{Khostovan25}, 3 from \citet{Jin19}, 3 from \citet{Brisbin17}, 2 from \citet{Daddi21}, 1 from \citet{Hasinger18}, 1 from \citet{casey15}, 1 from \citet{Ikeda25}, 1 from \citet{Sillassen25}, 1 from \citet{Jin2022}, 1 from \citet{Jin24b}, 1 from Jin (private communication). We also obtained 26 photometric redshifts from the COSMOS2020 catalog \citep{COSMOS2020}, 17 from the COSMOS2025 \citep[COSMOS-Web,][]{Shuntov25}, 3 from \citet{A3COSMOS}, 1 from \citet{Miettinen17} and,  1 from \citet{Vlugt23}.  The full procedure will be described in Carvajal-Bohorquez et al. (in prep.).

We identified counterparts for 80 2-mm sources (10 sources have no proxy detections), of which 18 have multiple counterparts. These 18 multiple sources break into 43 galaxies, including 19 with photometric and 24 with spectroscopic redshifts. Among the remaining 62 sources with a single counterpart, we have 29, 32 and 1 galaxies with photometric, spectroscopic, and no redshift, respectively. Finally, 11 galaxies are left without redshift (1 galaxy identified through a proxy but without redshift, and 10 2-mm sources without proxy). Compared to GOODS-N, the redshift identification completeness is lower, which limits the interpretation of redshift-dependent statistics.

\subsection{The SIDES simulations}

The simulated infrared dusty extragalactic sky (SIDES, \citealt{Bethermin2017,Bethermin2022}) is a semi-empirical realistic simulation of the sky from the mid-infrared to the millimeter. It is based on a dark-matter halo catalog from N-body numerical simulations, projected into a lightcone. The stellar mass of the galaxies hosted by the (sub-)halos is derived using the abundance matching technique (see, e.g., \citealt{Behroozi2010} or  \citealt{Moster2013} for a description of the method), which reproduces the observed stellar mass function by construction. The star-forming properties are then drawn following empirical scaling relations including their scatter, such as the relation between the star formation rate (SFR) and the stellar mass \citep{Schreiber2015}. The galaxy population on this relation is tagged as main sequence and the positive outliers as starbursts.

A spectral energy distribution (SED) template is then attributed to each galaxy depending if it is a main-sequence or starbursting object. These templates are based on observations, and follow the observed evolution of the mean radiation field $\langle U \rangle$ with redshift \citep{Magdis2012b,Bethermin2015}. The $\langle U \rangle$ parameter is related to the dust temperature (approximately $T_{\rm dust} \propto \langle U \rangle^\frac{1}{4+\beta}$, where $\beta$ is the dust spectral index), and is drawn assuming an observationally-motivated 0.2\,dex scatter around the mean evolution. Finally, the SEDs are integrated in broad-band filters. To evaluate properly the impact of cosmic variance on our observations, we use the new 117\,deg$^2$ version of SIDES introduced in \citet[][hereafter SIDES-Uchuu]{Gkogkou2023}.

SIDES includes the free-free and the synchrotron emission of star-forming galaxies in their SED template assuming the infrared-radio correlation of \citet{Sargent2010c} and an equality between synchrotron and free-free emission at 30\,GHz. The radio galaxies are not included in SIDES. They are discussed in Sect.\,\ref{sect:2mmonly}.

In \citet{Bing2023}, we already showed that SIDES reproduces well the number counts at both 1.2 and 2\,mm from single-dish and interferometric surveys, if we take into account resolution effects. In this paper, we will compare SIDES with measured colors and redshift distributions (see Sect.\,\ref{sect:properties}) to further validate it in the millimeter regime, before using it to discuss the relevance of 2\,mm surveys (see Sect.\,\ref{sect:discussion}).

\subsection{End-to-end N2CLS simulation based on SIDES}

\label{sect:e2e}

The source fluxes measured by single-dish millimeter surveys are affected by a large set of effects such as blending, data filtering, correlated noise, or flux boosting. To simulate only the blending effect, we generated noiseless maps using a Gaussian beam for the sources and we extracted the sources to produce the so-called "blob" catalog.

To include the other effects, we produced end-to-end (hereafter, E2E) simulations based on SIDES and real N2CLS data. The GOODS-N simulations are described in \citet{Bing2023}. For COSMOS, we generated a new simulation for the full dataset (see Carvajal-Bohorquez et al. in prep.) using a similar method. For each detector in the arrays, the incident flux is computed as a function of time using the SIDES simulated map and the real pointing coordinates, which vary with time. The resulting detector timelines are then combined with the real detector timelines after reversing the sign of half of them. These new timelines are used to produce simulated maps using PIIC, in which the true sky signal cancels out while the SIDES signal and the true noise remain. We then run the same source extraction procedure as described in Sect.\,\ref{sect:goodsn_cat} and \ref{sect:cosmos_cat}. A similar procedure is applied for each of the 117 SIDES-Uchuu 1-deg$^2$ tiles. For the need of this paper, we matched the simulated 2\,mm catalog with the 1.2\,mm one using the same procedure as done for the observational catalogs. The result is called hereafter E2E catalog.

We checked that the number of robust 2\,mm sources per simulated COSMOS field (71$\pm$16) is compatible at the 1.2-$\sigma$ level with the N2CLS (90). \citet{Bing2023} also found marginally higher number counts in their shallower COSMOS data than in GOODS-N and SIDES E2E. A similar result was found by \citet{Adscheid2024} using the A3COSMOS archival survey.

\section{Colors and redshifts of the 2\,mm sources}

\label{sect:properties}

\subsection{2\,mm versus 1.2\,mm colors}

\label{sect:colors}

\begin{figure}
\centering
\includegraphics[width=9cm]{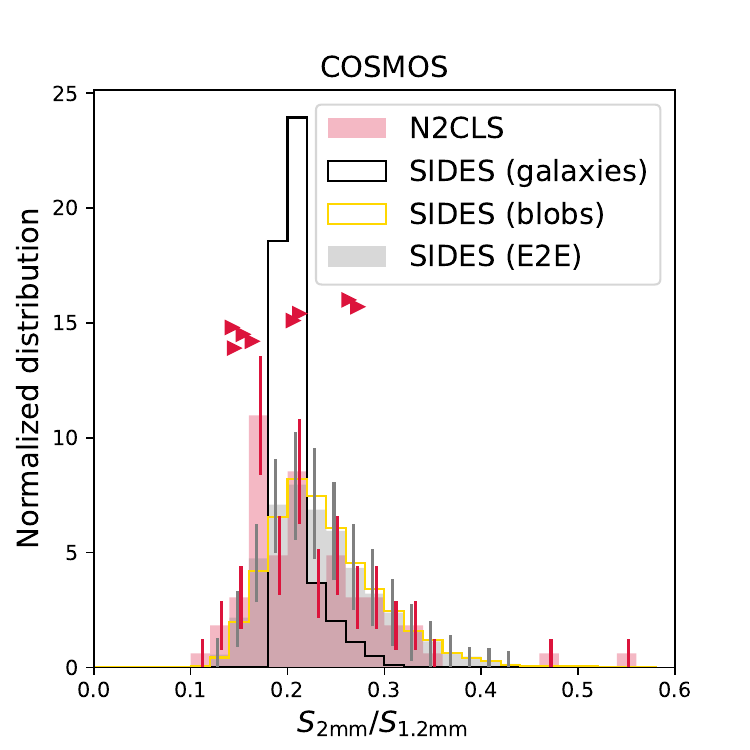}
\caption{\label{fig:col_histo} Distribution of the 2 to 1.2\,mm color for the true N2CLS COSMOS catalog (red filled histogram, only S/N$\ge4.6$ sources at both wavelengths and various simulated catalogs based on SIDES. The solid black histogram corresponds to the simulated SIDES galaxy catalog after applying a flux cut similar to COSMOS data (see Sect.\,\ref{sect:colors}). The yellow histogram is based on a similar selection, but applied to the blob catalog extracted from the noiseless simulated map. The grey histogram is the distribution obtained using the E2E simulation described in Sect.\,\ref{sect:e2e}. The error bars on the N2CLS histogram are computed assuming a Poisson law, and those from the E2E are the standard deviation between the various simulated fields. The histograms are normalized to have a unity area ($\int n(c)\, dc = 1$, where $c$ is the color). The lower limits on the color of 1.2\,mm non-detections are shown as right-pointing triangles. Their y-axis position is arbitrary and was chosen to reduce overcrowding in the figure.}
\end{figure}

\begin{table*}
\caption{\label{tab:colors} Comparison between the main characteristics (mean, median, standard deviations) of the 2 to 1.2\,mm color distributions of various catalogs. The N2CLS COSMOS catalog is the true catalog after selecting the high-reliability sources at both 1.2\,mm and 2\,mm (S/N$\ge$4.6). The SIDES galaxy and blob catalogs are both cut at the flux limit of the N2CLS COSMOS field (see Sect.\,\ref{sect:colors}). The SIDES E2E catalog is described in Sect.\,\ref{sect:e2e}. For the true N2CLS data, the uncertainties on the mean, the median, and the standard deviation are derived using a bootstrap method. For the simulated data, we provide the standard deviation between the multiple COSMOS-like fields extracted from the full SIDES-Uchuu simulation. The impact of the photometric calibration uncertainties on the colors ($\sim$7\,\%, see Sect.\,\ref{sect:colors}) are not taken into account in this table. The last column presents the probability associated with the KS test between a given distribution and the N2CLS.}
\centering
\begin{tabular}{lllll}
\hline
\hline
Catalog & Mean color & Median color & Standard deviation & KS p-value against N2CLS \\
\hline
N2CLS COSMOS & 0.222$\pm$0.008 & 0.206$\pm$0.007 & 0.070$\pm$0.010 & \\ 
SIDES galaxies (COSMOS flux cuts) & 0.212$\pm$0.003 & 0.205$\pm$0.003 & 0.022$\pm$0.003 & $2.2 \times 10^{-11}$\\ 
SIDES blobs (COSMOS flux cuts) & 0.240$\pm$0.002 & 0.229$\pm$0.002 & 0.062$\pm$0.004 & $1.9 \times 10^{-3}$\\ 
SIDES E2E (COSMOS-like) & 0.238$\pm$0.010 & 0.227$\pm$0.010 & 0.063$\pm$0.012 & $6.9 \times 10^{-3}$\\
\hline
\end{tabular}
\end{table*}

Thanks to its simultaneous observations at 1.2\,mm and 2\,mm, the N2CLS is ideal to study the millimeter colors of extragalactic sources. We focus our analysis on the COSMOS field for which we have a significantly larger number of detections. Since the color is a ratio, it is affected by the uncertainties on both the numerator and the denominator. We thus limit our analysis to high-reliability (S/N$\ge4.6$) sources at both 1.2\,mm and 2\,mm. However, in COSMOS, this choice has no impact compared to a S/N$\ge3.9$ cut at 1.2\,mm (80\,\% purity), since there are no robust 2\,mm sources with an S/N between 3.9 and 4.6 at 1.2\,mm. The measured 2\,mm versus 1.2\,mm color distribution is shown in Fig.\,\ref{fig:col_histo}. This distribution is asymmetric with a long tail of high color values, corresponding to red objects.

To interpret these data, we compute several distributions based on the SIDES simulation. In this work, we use the 117\,deg$^2$ version of the simulation \citep{Gkogkou2023} up to z$\sim$7 to estimate the field-to-field variance. We estimated the impact of this redshift limit using the older 2\,deg$^2$ version with redshifts up to z=10 \citep{Bethermin2017}, and found that it is negligible. For instance, the mean color varies by only 0.3\,\%. For the individual galaxy color distribution, we just cut the SIDES simulated galaxy catalog at the typical flux cut of the N2CLS catalog. For this purpose, we use the S/N limit of 4.6 and multiply it by the mean 1-$\sigma$ depth of our maps, and obtained the following selection: S$_{\rm 2\,mm} \ge 0.414$\,mJy and S$_{\rm 1.2\,mm} \ge1.38$\,mJy. To evaluate the impact of the NIKA2 angular resolution, we applied similar cuts to the blob catalog presented in Sect.\,\ref{sect:e2e}. Finally, we use the catalogs resulting from our E2E simulations (including real noise in the simulated timelines, the instrumental beam, and map-making effects) also introduced in Sect.\,\ref{sect:e2e} to compare accurately SIDES to N2CLS. Since these simulations produces 117\,independent fields with the same field geometry as N2CLS COSMOS, we can use it to estimate the field-to-field variance. These uncertainties are larger than the ones obtained from a bootstrap method, since galaxies in a given region of the sky tend to accumulate at similar redshift and thus have similar colors. These correlations between sources increase the variations between fields of the various summary statistics. The three distributions obtained from SIDES are presented in Fig.\,\ref{fig:col_histo}.

The means and medians of all these distributions are listed in Table\,\ref{tab:colors}, and are roughly compatible. The SIDES galaxy catalog have a slightly lower mean but the same median as N2CLS, while the blob and E2E catalogs have higher means and medians. This shows that the resolution effects have a small impact ($\sim$10\,\%) on the average colors. To evaluate the quality of SIDES predictions, the most relevant comparison is between the N2CLS and the E2E, since they should be affected by exactly the same instrumental and source extraction effects. The difference between the two means and medians corresponds to 1.2 and 1.7 times the field-to-field standard deviation. We also have to take into account the calibration uncertainties. Because the 1.2 and 2\,mm fluxes are affected in the same way (i.e. systematic uncertainty), the absolute calibration uncertainty does not have to be taken into account for the color uncertainty. On the other hand, the point-source rms calibration uncertainties (6\% at 1.2\,mm and about 3\% at 2\,mm, \citealt{Perotto2020}), lead to a 6.7\,\% statistical uncertainty on the color.
By combining quadratically these uncertainties with the field-to-field variance, we obtain uncertainties of 0.020 and 0.018 on the mean and the median color, corresponding for both to a 0.8 and 1.1\,$\sigma$ difference. There is thus no significant tension on the average colors of the N2CLS sources and the SIDES predictions.

However, the distributions showed in Fig.\,\ref{fig:col_histo} exhibit some difference of shapes. The most striking one is how narrow is the SIDES galaxy color distribution compared to the others with a standard deviation which is $\sim$3 times smaller. This is not the case for the SIDES blob catalog, suggesting that the resolution effects drive the dispersion in color rather than the noise. In contrast, the N2CLS and the SIDES E2E catalogs agree on the dispersion at the 0.5\,$\sigma$ level, demonstrating how efficient our E2E simulation is to predict the observations.

Despite the agreement between N2CLS and SIDES E2E on these basic metrics, the distributions are not exactly similar. For instance, the probabilities returned by the Kolmogorov-Smirnoff (KS) null-hypothesis test comparing N2CLS and SIDES are always lower than 0.01. Above a p-value of 0.05, we can consider that the two distributions are drawn from the same probability density function. In contrast, a low probability indicates that the two distributions are not similar. This test does not take into account the field-to-field variance and the calibration uncertainties. It may thus be improper, and must be interpreted with caution. However, there is a 2.5-$\sigma$ excess in the 0.16--0.18 bin after taking into account the field-to-field variance. If we exclude this color bin from the KS test, we obtain a p-value of 0.23, confirming that the disagreement is caused by this single bin. 
The origin of this peak in the color distribution remains unclear.

Finally, we also checked the mean color in the GOODS-N field. This field is limited by its small number of 2\,mm detection (25), but it is deeper. We find a mean color of 0.196$\pm$0.014, which is compatible with the 0.208 value found in the SIDES galaxy catalog after applying flux cuts similar to GOODS-N (S$_{\rm 2\,mm} \ge 0.2$\,mJy and S$_{\rm 1.2\,mm}\ge0.71$\,mJy. Contrary to COSMOS, we do not observe any excess in the 0.16--0.18 color bin, suggesting that this feature is a field variance effect.

\subsection{Redshift distribution}
 
\label{sect:Nz}

\begin{figure}
\centering
\includegraphics[width=9cm]{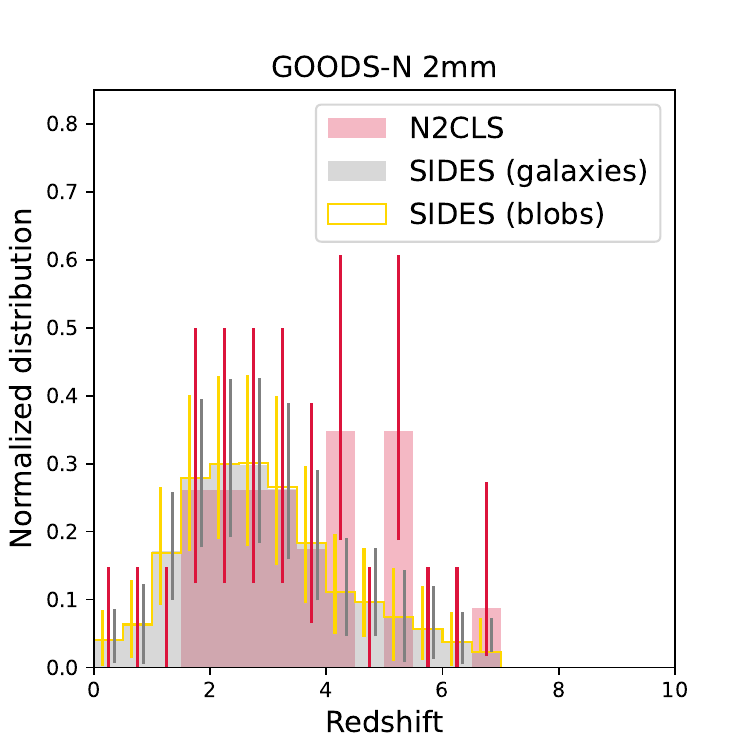}
\caption{\label{fig:Nz_GOODSN} Redshift distribution of the GOODS-N high-reliability 2\,mm sources. The red histogram is the distribution measured in the N2CLS. When sources are multiple, the redshift of the main component is used (see Sect.\,\ref{sect:2mmonly}). The uncertainties correspond to a Poisson law. The grey histogram is derived from the SIDES galaxy catalog after applying a flux cut similar to the GOODS-N field (see Sect.\,\ref{sect:Nz}). The uncertainties correspond to the field-to-field variance of a 15.9$\times$10\,arcmin$^2$ field. The yellow histogram is based on a similar selection, but applied to the SIDES blob catalog and using the redshift of the brightest galaxy in the beam. The histograms are normalized to have a unity area ($\int n(z)\, dz = 1$).}
\end{figure}

\begin{figure}
\centering
\includegraphics[width=9cm]{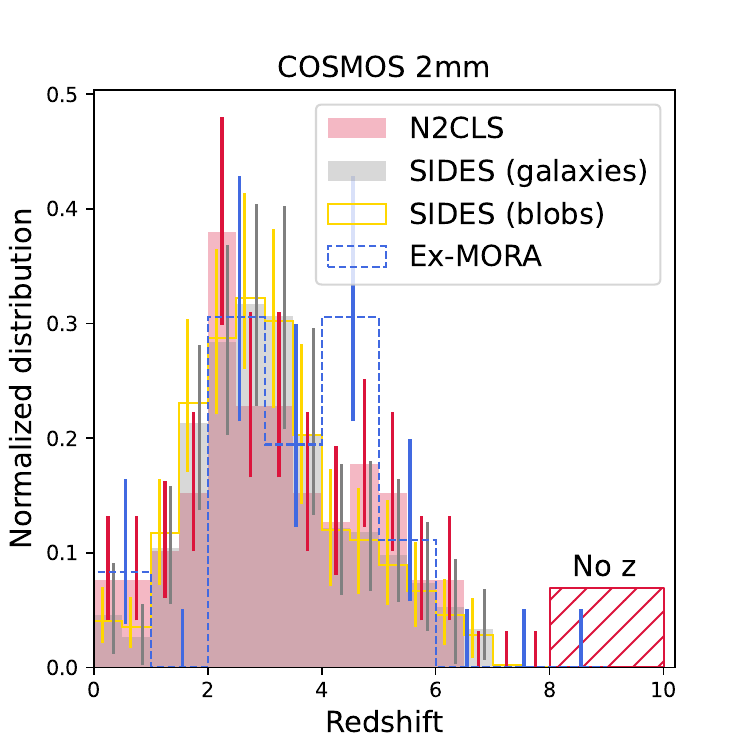}
\caption{\label{fig:Nz_COSMOS} Same as Fig.\,\ref{fig:Nz_GOODSN}, but for the COSMOS field. The blue dashed histogram represents the Ex-MORA survey \citep{Long2024}.}
\end{figure}

The redshift distribution is an important constraint for models. For a flux-limited sample selected at a single wavelength, this distribution varies with both the flux cut and the wavelength \citep[e.g.,][]{Smolcic2012,Zavala2014,Bethermin2015b}. In the (sub)millimeter, the mean redshift is expected to increase with the wavelength. With its combination of size and depth, the N2CLS 2\,mm survey is ideal to test model predictions in the long-wavelength and faint-flux regime. Since it has a complete redshift association, we focus first our analysis on the GOODS-N field in which all NIKA2 sources have at least one associated galaxy with a known redshift \citep{Berta2025}.

In Fig.\,\ref{fig:Nz_GOODSN}, we present the redshift distribution of 2\,mm high-reliability sources with Poisson uncertainties. In case of a multi-component source, we use the redshift of the main component (listed as "a" in \citealt{Berta2025})\footnote{The source N2GN\_1\_34 main component has no redshift, and is thus not included in our analysis.}. We compare these results with the predictions from SIDES. The source identification is a complex process involving multi-wavelength data. 

Since SIDES does not produce simulated optical and radio images, we had to rely on two simplified approaches. First, we selected a sample in SIDES by cutting the galaxy catalog above the survey flux limit (S$_{\rm 2\,mm} \ge S/N_{\rm lim} \times \sigma = 0.2$\,mJy). This sharp flux limit is a simplification, since the survey depth varies across the field and the transfer function can also alter the true flux limit. However, the redshift distribution is only weakly dependent on the exact flux cut \citep{Bethermin2015b}, so an approximate flux limit is sufficient for our purpose. Our second approach takes resolution effects into accounts and is based on the SIDES blob catalog. There are always several galaxies with different redshifts present in the same NIKA2 beam, and we attribute the redshift of the brightest galaxy at 2\,mm to the blob (see \citealt{Bethermin2017,Bethermin2024} for a detailed presentation of the method). For both approaches, we compute the field-to-field variance by dividing the SIDES catalog into 15.9$\times$10\,arcmin$^2$ subfields (2\,541 in total). The galaxy (gray histogram) and blob (yellow histogram) are very similar (see Fig.\,\ref{fig:Nz_GOODSN} and Fig.\,\ref{fig:Nz_COSMOS}). This shows that, contrary to the number counts, the redshift distribution is not sensitive to the resolution effects \citep[see, e.g.,][]{Bing2023}.

There is an overlap between the 1\,$\sigma$ error bars of the N2CLS/GOODS-N and both galaxy and blob SIDES redshift distributions in all the redshift bins except the $5.0<z<5.5$ one, where N2CLS is $\sim$1.5\,$\sigma$ above the SIDES predictions. This is due to a known galaxy overdensity at z$\sim$5.2 around the HDF850.1 galaxy. A companion N2CLS paper shows that this overdensity contain numerous dusty star-forming galaxies \citep{Lagache2025}. There is also an excess in the $4.0<z<4.5$ bin, but the 1-$\sigma$ error bars overlap and it is thus not significant. We measure a mean redshift of 3.6$\pm$0.3, while the SIDES simulation predicts 2.9$\pm$0.2 for both the galaxy and blob selection. The redshift limit at z$\sim$7 has a minor impact on this quantity and we found a mean redshift of 3.0 using the 2\,deg$^2$ version of SIDES which is cut at z=10 \citep{Bethermin2017}\footnote{The number of objects as a function of redshift in SIDES exhibit a steady and strong decrease above z=4. The number of S$_{\rm 2\,mm}\ge$0.2\,mJy at z$>$7 is 23, while it is only 3 at z$>$8 and 0 at z$>$9. It is thus reasonable to assume that there would be no detectable 2\,mm source at z$>$10 if the simulation was extended to higher redshift.}. This $\sim$2\,$\sigma$ difference is likely due to the overdensity at z$\sim$5.2.

To test this, we performed a KS test on both redshift distributions after excluding the $5.0<z<5.5$ range. The test returns a p-value of 0.4 (same values for the SIDES galaxy and blob catalogs), and N2CLS in GOODS-N and SIDES are thus fully compatible after excluding this known overdensity, while it is only marginally compatible (p-value of 0.06 for both the galaxy and blob catalogs) if we include this redshift bin.

The COSMOS 2\,mm catalog is slightly incomplete in redshift (only 88\,\% have a main component with an associated redshift), but has a significantly larger number of detections. We thus performed a similar analysis as in GOODS-N, but using a S$_{\rm 2\,mm}$ flux cut of 0.414\,mJy for the SIDES selection and a 34$\times$34\,arcmin$^2$ for the field-to-field variance. This field size matches the 1130\,deg$^2$ area where sources are extracted in the real N2CLS data (Carvajal-Bohorquez et al. in prep). However, the results should be taken with caution because of the mild redshift incompleteness. The result is shown in Fig.\,\ref{fig:Nz_COSMOS}. We obtain a mean N2CLS redshift of 3.2$\pm$0.2, which agrees with SIDES galaxy prediction of 3.2$\pm$0.2 and the blob prediction of 3.1$\pm$0.1. The 1\,$\sigma$ error bars of the N2CLS COSMOS and the SIDES distribution overlap, and the KS test returns a p-value of 0.58 for the galaxy catalog and 0.47 for the blob catalog, compatible with the two samples being drawn from the same distribution. 

We also compare our N2CLS results with those of the Ex-MORA survey \citep{Long2024}, which has a similar depth and a partial overlap. The redshift distributions of the two surveys agree all the 1-$\sigma$ error bars overlapping except in the 1$<$z$<$2 bin, where this is almost the case. The KS test p-value is 0.2, confirming the agreement between the two surveys.

Contrary to GOODS-N, there is no tension between the observed and predicted mean redshifts. This suggests that GOODS-N is a peculiar field with a strong overdensity at z$\sim$5.2. However, COSMOS is slightly incomplete, and could be more incomplete at higher redshift, biasing the mean redshift towards lower values. 

\subsection{Colors versus redshift}

\label{sect:col_vs_z}

\begin{figure*}
\centering
\begin{tabular}{cc}
\includegraphics[width=9cm]{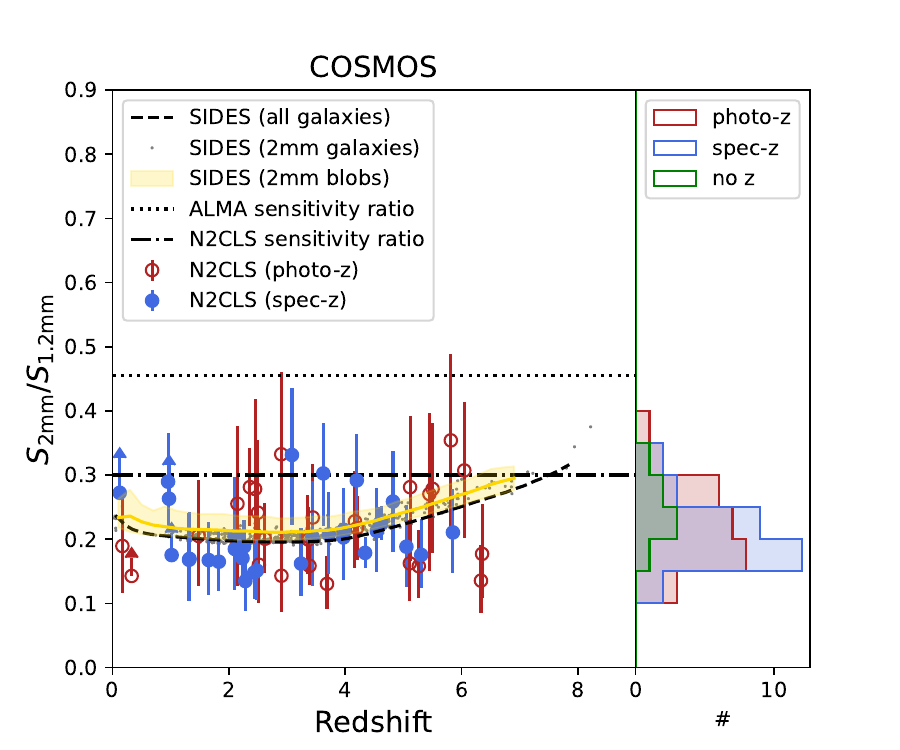} & \includegraphics[width=9cm]{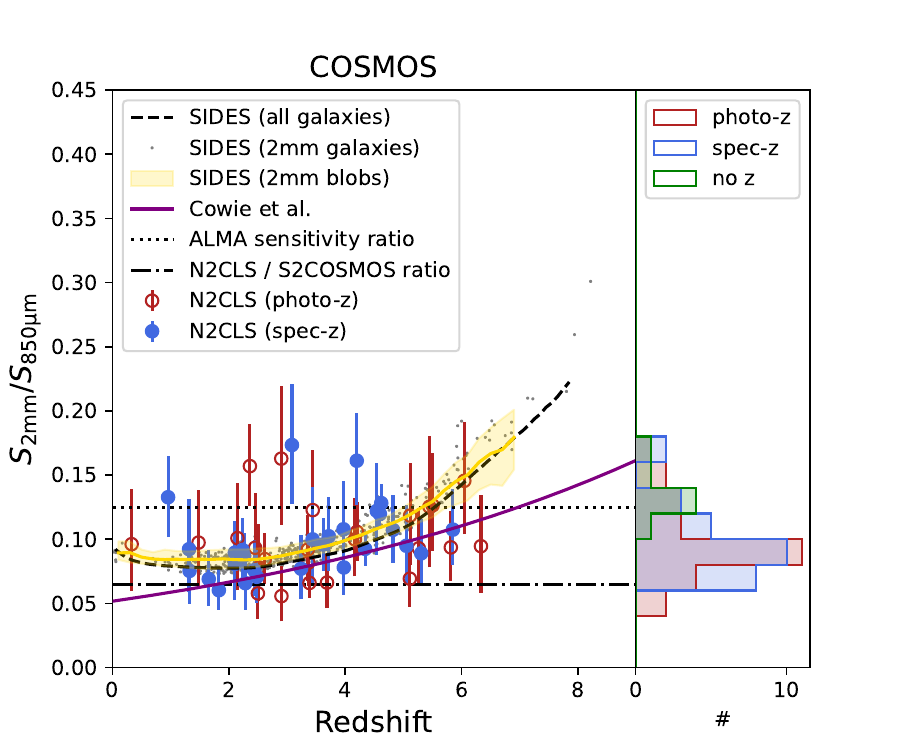}\\
\includegraphics[width=9cm]{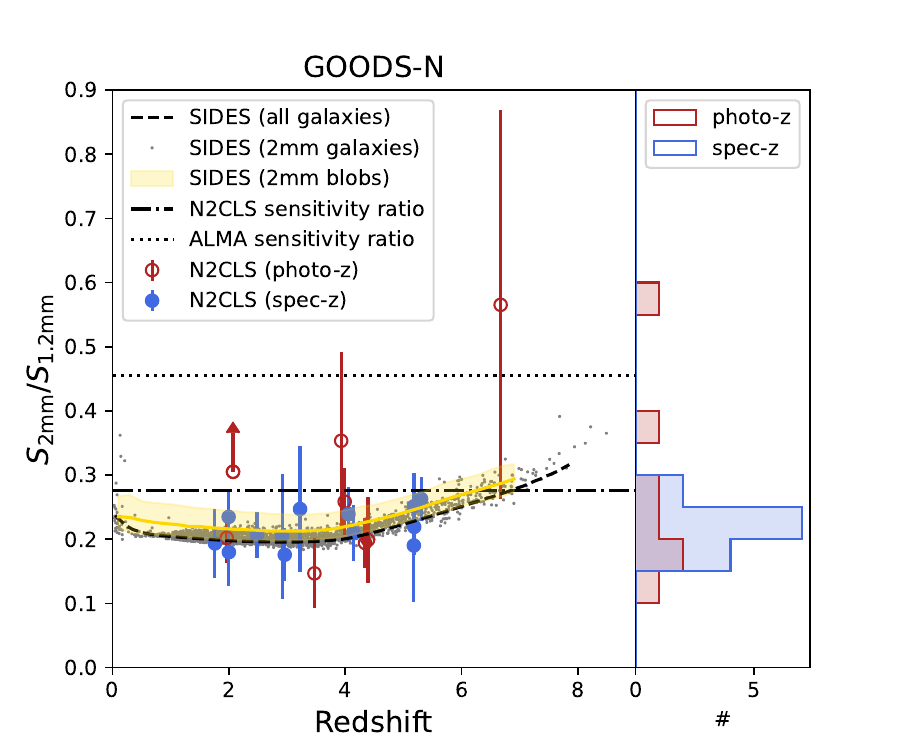} & \includegraphics[width=9cm]{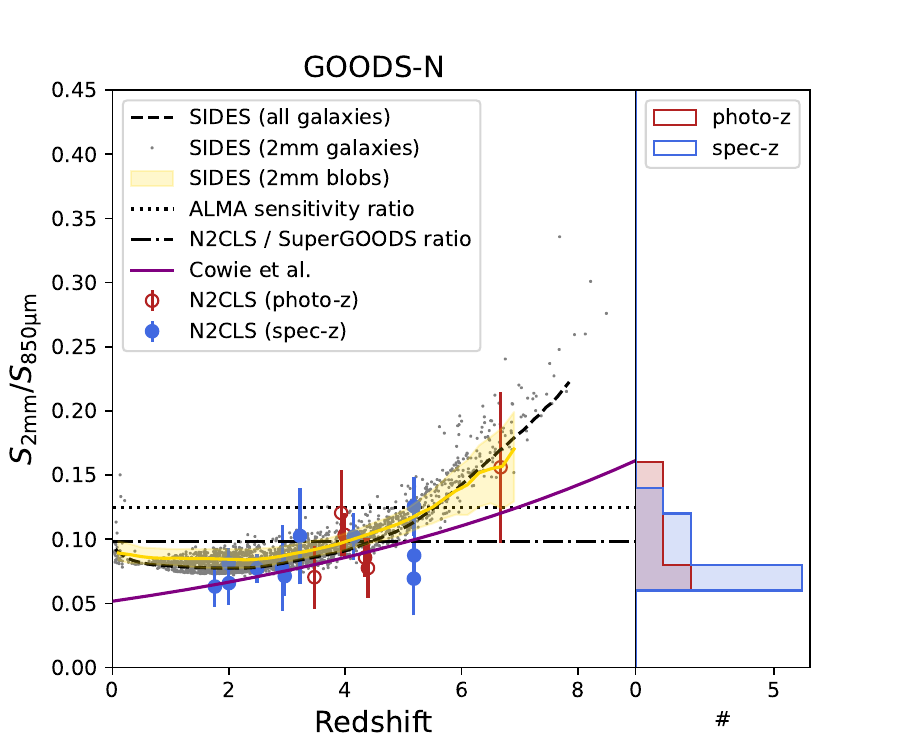}\\
\end{tabular}
\caption{\label{fig:col_vs_z} Color as a function of the redshift in the COSMOS (top) and GOODS-N (bottom) fields. The left plots show the internal NIKA2 color (S$_{\rm 2\,mm}$/S$_{\rm 1.2\,mm}$) and the right ones show the ratio between the N2CLS 2\,mm  and the SCUBA2 850\,$\mu$m fluxes from S2COSMOS \citep{Simpson2017} and SUPER GOODS \citep{Cowie2017}. The filled blue circle are the N2CLS sources with spectroscopic redshifts, while the open red circles corresponds to photometric redshifts. The grey dots are from the SIDES 2\,deg$^2$ simulated catalog after applying a 2\,mm flux selection similar to SIDES (see Sect.\,\ref{sect:Nz}). The yellow solid line corresponds to the median of the SIDES blob catalog, using the redshift of the brightest galaxy in the beam, and the colored area correspond to the 1\,$\sigma$ confidence region. The dashed line is the mean color as a function of redshift from SIDES 2\,deg$^2$ without applying any selection. We also show the sensitivity ratio ($\sigma_{\rm 2\,mm} / \sigma_{\rm 1.2\,mm}$ where $\sigma_{\rm 2\,mm}$ and $\sigma_{\rm 1.2\,mm}$ are the RMS of the noise at 2\,mm and 1.2\,mm, respectively) for ALMA (dotted line) and N2CLS (dot-dashed line) ; see the discussion in Sect.\,\ref{sect:disc_obs}. The right-side panel of each plot show the color distribution of sources with a photometric redshift (red), a spectroscopic redshift (blue), and no redshift (green).}
\end{figure*}

Since the dust emission of galaxies peaks at $\sim$80\,$\mu$m rest-frame, red colors in the (sub-)millimeter can be used to find high-redshift candidates. For instance, \textit{Herschel} surveys identified sources with red \textit{Herschel}/SPIRE colors (usually 250 versus 500\,$\mu$m) as z$\gtrsim$4 candidates \citep{Dowell2014,Asboth2016,Donevski2018}. However, \citet{Bethermin2017} and \citet{Donevski2018} showed using simulations that a large fraction of these sources have redder observed colors than their intrinsic ones because of fluctuations of the noise in the bands used to compute the color (positive in the long-wavelength band and negative in the short-wavelength one). Similarly, millimeter colors were proposed to identify candidates at even higher redshifts \citep[e.g.][]{Casey2020,Cooper2022,Cowie2023}. In this section, we study the (sub-)millimeter color of N2CLS galaxies versus redshift to test this approach.

\subsubsection{Colors versus redshift in the N2CLS}

In the color-versus-redshift relation, we can only place sources which have all their flux coming from the same redshift. Thus, we selected only galaxies with a single counterpart to have no ambiguity. 
We derived both the N2CLS colors (S$_{\rm 2\,mm}$/S$_{\rm 1.2\,mm}$) and the 2\,mm versus 850\,$\mu$m colors after matching our catalogs with the S2COSMOS \citep[]{Simpson2017} and SUPER GOODS \citep{Cowie2017} SCUBA2 catalogs in COSMOS and GOODS-N, respectively. Contrary to Sect.\,\ref{sect:colors}, we use the full 1.2\,mm catalog to compute the colors. The results are presented in Fig.\,\ref{fig:col_vs_z}.

There is no clear trend of the observed N2CLS S$_{\rm 2\,mm}$/S$_{\rm 1.2\,mm}$ color versus redshift (Fig.\,\ref{fig:col_vs_z}, left), and data points exhibit a large scatter. For a linear fit of the color versus redshift, we find a slope of $0.009 \pm 0.005$ when combining GOODS-N and COSMOS. The color measurements are affected by the uncertainties in both the numerator and the denominator. Since we account for both the limited signal-to-noise ratio and the uncertainties in the deboosting, this can lead to large relative uncertainties ($>$1/3), and the N2CLS sample is thus too small to detect any  significant trend.

The observed evolution of the 2\,mm versus 850\,$\mu$m color with redshift exhibits a tentative mild increasing trend (Fig.\,\ref{fig:col_vs_z}, left) with a slope of 0.014$\pm$0.006 in COSMOS corresponding to a 2.3\,$\sigma$ significance. This result agrees with \citet{Cooper2022}, \citet{Cowie2023}, and \citet{Long2024}, who also found a weak trend without indicating its significance. We attempted to fit separately the trend below and above z=2.5 (motivated by the trend predicted by SIDES in Sect.\,\ref{sect:sides_col_vs_z}). The results are not statistically significant (0.002$\pm$0.026 at z$<$2.5 and 0.017$\pm$0.011 at z$>$2.5), but they are compatible with the flat trend below z=2.5 and the rise above predicted by SIDES.

In the COSMOS field, a small fraction of the sources has no redshift (12\,\%). We compared the color distributions of the sources depending of the type of information on their redshift (see the right side panels of each sub-figure of Fig.\,\ref{fig:col_vs_z}). For the N2CLS S$_{\rm 2\,mm}$/S$_{\rm 1.2\,mm}$ colors in COSMOS, the three distributions are similar, except a mild tension between the sources with a spectroscopic redshift and without redshift in COSMOS (p-value of 0.04 to have similar parent distributions according to the KS test). The redshift incompleteness of our sample should thus not affect significantly our results. Concerning the S$_{\rm 2\,mm}$/S$_{\rm 850\,\mu m}$ colors in COSMOS, there is a tension between both the photometric and spectroscopic redshift samples and the objects without redshift (p-value of 0.005 and 0.003, respectively). There is no tension between the other distributions. The sources without redshift tend to be redder. It is thus possible that the sample of 11 COSMOS 2\,mm galaxies without redshift (no proxy or no redshift available for the counterpart) is biased towards cold or high-z sources.

\subsubsection{Colors versus redshift in SIDES}

\label{sect:sides_col_vs_z}

We compare our results with the intrinsic colors of the galaxies in SIDES after applying the same 2\,mm flux cut as in Sect.\,\ref{sect:colors}. We use the 2\,deg$^2$ version of the simulation \citep{Bethermin2017} for this analysis, since it goes up to z$\sim$10 and is wide enough for this qualitative comparison. The results are shown in Fig.\,\ref{fig:col_vs_z} as grey filled circles. We remark that the intrinsic scatter predicted by SIDES is much smaller than the observational scatter, which is driven by measurement uncertainties. We also computed the mean color of the full SIDES galaxy catalog without applying any flux selection (dashed line). Finally, we derived the evolution of the color in the SIDES blob catalog as a function of the redshift of the brightest component (yellow line and shaded area).

The SIDES simulation includes only a scatter on the mean intensity of the radiation field $\langle U \rangle$, which is directly connected to the dust temperature. In contrast, the slopes of the SEDs in the Rayleigh-Jeans regime have only small variations. Since most of the sources are observed close from this regime, the S$_{\rm 2\,mm}$/S$_{\rm 1.2\,mm}$ color at fixed redshift has thus only a very small scatter. We could thus think that SIDES may severely under-predict the color dispersion. However, as shown in Sect.\,\ref{sect:colors} and Fig.\,\ref{fig:col_histo}, the observed color scatter is 1-$\sigma$ compatible with the SIDES scatter after applying all the observational effects. There is thus no evidence that the SIDES color scatter is underestimated.

The mean NIKA2 color in both SIDES blob and galaxy catalogs is nearly constant ($\sim$0.2) between z=0.5 and 4. It increases slightly at higher redshifts ($\sim$0.3 at z=8). The colors obtained for the 2\,mm-selected SIDES galaxy (gray symbols) are redder (higher values) than the mean value of all individual galaxies in the SIDES simulation (dashed line). This selection bias will be discussed in Sect.\,\ref{sect:disc_model}. The color slightly increases at z$<$0.5 (as shown by the dashed line for the SIDES mean and the gray dots for individual 2\,mm-selected SIDES galaxies) due to the free-free and synchrotron contributions included in the SIDES templates \citep{Bethermin2012c}. Additionally, we note that the blob selection produces slightly redder colors at low redshifts compared to the 2\,mm-selected galaxy catalog.

The results are similar for the S$_{\rm 2\,mm}$/S$_{\rm 850\,\mu m}$ color (Fig.\,\ref{fig:col_vs_z}, right). However, the reddening of the colors with increasing redshift starts at lower redshift (z$\sim$2.5). This is expected since the Rayleigh-Jeans power-law regime starts above 200\,$\mu$m rest-frame. Below this rest-frame wavelength, the SED is flatter and this regime is thus reached at lower redshift for the observed 850\,$\mu$m than the 1.2\,mm. The blob catalog produces marginally redder colors at low z and bluer ones at z$>$6 than the 2\,mm-selected galaxy catalog. \citet{Cowie2023} proposed a power-law relation between the logarithm of the color and the redshift (log$_{10}(S_{\rm 2\,mm}/S_{\rm 850\,\mu m}) = 0.055 z - 1.287$), which disagrees with SIDES forecast at z$<$2 and z$>$5. Unfortunately, we lack data in these redshift ranges to test if SIDES is correct or if the power law is a good approximation.

\subsection{2\,mm sources without 1.2\,mm counterparts}

\label{sect:2mmonly}

Because of the expected mild evolution of the NIKA2 color with redshift at z$\gtrsim$4, the sources detected only at 2\,mm could be potentially at very high redshift. However, our sample has a 95\,\% purity and these sources could thus be spurious.

In the GOODS-N field, only one source (N2GN\_2\_13) is detected only at 2\,mm. As explained in \citet{Berta2025}, this source was detected by NOEMA and has a short-wavelength counterpart with a photometric redshift of 2.071. This suggests that their is no spurious source in the GOODS-N sample. Based on the 25 detections and the 5\,\% spurious rate, we expect in average 1.25 spurious sources per GOODS-N-sized field, and the Poisson probability of having 0 spurious objects in such a field is 0.29. This absence of spurious source is thus compatible with our estimated sample purity.

In the COSMOS field, out of the 90 high-reliability 2\,mm sources, we found 8 sources without 1.2\,mm counterpart (S/N$\ge$3.9 cut). The two brightest 2\,mm-only sources (N2CO\_2\_37 and N2CO\_2\_44) are associated (matching distance of 1.40 and 1.42\,arcsec, respectively) with two radio sources (COSMOSVLADP\_J100114.85$+$020208.8 and COSMOSVLADP\_J095945.19$+$023439.3, respectively) with spectroscopic redshifts of 0.97 and 0.12, respectively. N2CO\_2\_37 has a 4\,GHz flux density of 7.5\,mJy \citep{Schinnerer2010}. Extrapolating this emission at 150\,GHz (2\,mm) by assuming a standard synchrotron spectrum in $\nu^{-0.8}$, we obtain 0.41\,mJy (and 0.59\,mJy for $\nu^{-0.7}$). The majority of its 2\,mm N2CLS flux (0.71\,mJy) is thus expected to come from synchrotron emission. For the second source, which is more than one order of magnitude fainter in radio, the synchrotron is less likely to contribute, but we cannot exclude a contamination of dust emission by some free-free emission.

To understand the nature of the six remaining sources, we used the end-to-end simulation introduced in Sect.\,\ref{sect:e2e} to estimate the expected number of S/N$\ge$4.6 2\,mm sources without S/N$\ge3.9$ 1.2\,mm counterpart, and found 5.8$\pm$3.0 objects. The observations are thus fully compatible with our simulation. We also estimated the probability using another approach by computing the probability to find 6 or more spurious sources assuming a mean expected value of 4.5 (5\,\% spurious rate times 90 sources found in the N2CLS), and found a probability of 0.29. It is thus statistically possible that these 6 sources are all spurious. Searching for these 2\,mm-only sources is thus not a reliable way to select very high-z candidates.

\subsection{Comparison with the Ex-MORA survey}

\label{sect:MORA}

\begin{figure}
\centering
\includegraphics[width=9cm]{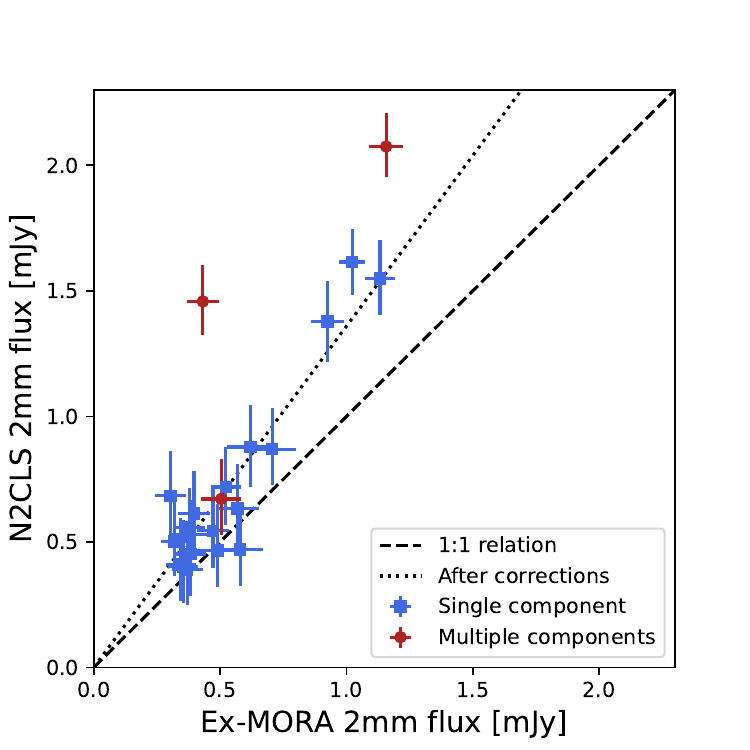}
\caption{\label{fig:comp_mora} Comparison between the exMORA (x axis, \citealt{Long2024}) and N2CLS (y axis) 2\,mm flux. The N2CLS sources with a single counterpart are in blue and the multiple sources are in red. The black dashed line indicates the one-to-one relation and the dotted line includes the various corrections (bandpass, source blending, and source sizes) discussed in Sect.\,\ref{sect:MORA}.}
\end{figure}

In the COSMOS field, the extended mapping obscuration to reionization with ALMA (Ex-MORA) survey (\citealt{Long2024}, see also \citealt{Casey2021} and \citealt{Zavala2021} for the first MORA survey) has produced an interferometric mosaic of 577\,arcmin$^2$ at 2\,mm, overlapping with the N2CLS footprint\footnote{The overlapping area was computed using MOCPy (\citealt{MOC,MOCPy}). We consider only the region where sources were extracted for the N2CLS.} on 533\,arcmin$^2$. In this section, we compare the Ex-MORA and N2CLS catalogs. The Ex-MORA survey is slightly deeper than N2CLS with flux limit of $\sim$0.3\,mJy versus $\sim$0.4\,mJy. Out of the 37 Ex-MORA sources, four of them are on the noisy edge of the N2CLS field and thus outside of the region where sources were extracted (eMORA.18, eMORA.30, eMORA.31, eMORA.36). Of the 33 remaining Ex-MORA sources, 22 have a corresponding N2CLS 2\,mm detection in the full catalog with S/N$\ge$3.9\footnote{We used a matching radius of 18\,arcsec, but all the counterparts are located at less than 6\,arcsec from the Ex-MORA source (1/3 of the beam FWHM) and there is no double matches.} For most of the sources without counterpart, we can observe a faint signal (1.5$\le$S/N$\le$3.9) in the N2CLS map at the Ex-MORA position: S/N=3.4 for eMORA.5, S/N=3.3 for eMORA.6, S/N=1.5 for eMORA.9, S/N=1.9 for eMORA.17, S/N=2.8 for eMORA.19, S/N=2.1 for eMORA.23, S/N=3.6 for eMORA.28, S/N=2.9 for eMORA.29, S/N=2.9 for eMORA.32, S/N=2.5 for eMORA.33. Only the source position of eMORA.34 has no positive fluctuation of the signal (S/N=-0.7), and is among the lowest S/N of the Ex-MORA survey (5.1).

We also investigated the Ex-MORA counterparts of robust (S/N$\ge$4.6) N2CLS 2\,mm sources. Of the 54 N2CLS sources within the Ex-MORA footprint, only 18 have corresponding Ex-MORA detections within the N2CLS beam ($<$9\,arcsec from the N2CLS centroid). In addition, we inspected the Ex-MORA maps and found at least one candidate source with an S/N$>$4 signal in the beam of 13 other N2CLS sources. Four other sources are located in a noisy region at the edge of the Ex-MORA survey, where no detection is possible. Thus, we are left with 19 sources without any significant counterparts in the NIKA2 beam. Only four of them are 2\,mm-only detections (see Sect.\,\ref{sect:2mmonly}) and are potentially spurious. The 15 others are also detected at 1.2\,mm and are likely true objects missed by Ex-MORA.

In Fig.\,\ref{fig:comp_mora}, we show the comparison between the N2CLS and the Ex-MORA 2\,mm flux of objects detected by both surveys. Most of the low-flux source (0.3--0.6\,mJy) have compatible fluxes at 2\,$\sigma$, but we can observe a small average excess of N2CLS flux compared to Ex-MORA. This excess is striking at higher flux, where the relative uncertainties are much smaller. A similar excess has been found in GOODS-N for four 2\,mm sources followed up by NOEMA \citep{Berta2025}. For sources with a single known component, the N2CLS fluxes have a mean excess ratio of 1.3, and it reaches 2.2 for multiple sources. In the case of multiple sources, the flux difference is straightforward to explain. Indeed, for our multiple sources, Ex-MORA is detecting only one of the counterparts, leading to a significantly smaller Ex-MORA flux than the total flux from the galaxies in the NIKA2 beam. 

The origin of the non-unity flux ratio between Ex-MORA and single-component N2CLS sources is less clear. It could be due to the combination of several effects. A first systematic factor comes from the large bandpass of NIKA2 and its different central observed wavelengths compared to Ex-MORA. We compute the flux density ratio between the two instruments assuming a power-law spectrum in $\lambda^{-2.95}$ (or $\nu^{2.95}$) based on the measured mean color between 1.2\,mm and 2\,mm (Sect.\,\ref{sect:colors} and Table\,\ref{tab:colors}):
\begin{equation}
\frac{S_{\rm 2\,mm}^{\rm NIKA2}}{S_{\rm 2\,mm}^{\rm Ex-MORA}} = \frac{\int \left ( \frac{\nu}{\nu_{\rm Ex-MORA}} \right )^{2.95} t_\nu \, d\nu}{\int t_\nu d\nu} = 1.085,
\end{equation}
where the central frequency of Ex-MORA is $\nu_{\rm Ex-MORA}$ = 147\,GHz and $t_\nu$ is the transmission of the NIKA2 2\,mm bandpass as a function of frequency. This ratio increase to 1.169 when assuming a modified blackbody in the Rayleigh-Jeans regime and with an emissivity index $\beta$ of 2.4 as found by \citet{Cooper2022}.

In addition, the flux measured in single-dish blobs can be significantly larger than the flux of the brightest galaxy in the beam \citep{Hayward2013,Karim2013,Cowley2015,Scudder2016,Bethermin2017,Bing2023} due to the presence of other galaxies in the beam. To evaluate the impact of this effect, we compared the flux in the SIDES blob catalog (Sect.\,\ref{sect:e2e}) with the flux of the brightest galaxy in the beam (see \citealt{Bethermin2024} for a detailed description of the method), and found a median ratio of 1.14.

Finally, the Ex-MORA flux is measured at the peak of emission, but this method can slightly underestimate the flux if the sources are marginally resolved (e.g., Sect.\,3.5 of \citealt{Bethermin2020}). Assuming Gaussian profiles for both the beam and the galaxy, the peak flux underestimates the total flux by the ratio between the synthesized beam area ($\Omega_{\rm beam} \propto a_{\rm beam} \,b_{\rm beam}$, where $a_{\rm beam}$ and $b_{\rm beam}$ are the beam FWHM of the major and minor axes) and the area of its convolution with the galaxy profile ($\Omega_{\rm beam} \propto \sqrt{(a_{\rm beam}^2 + s_{\rm intr}^2)\, (b_{\rm beam}^2 + s_{\rm intr}^2)}$, where $s_{\rm intr}$ is the intrinsic FWHM of the galaxy). For a typical Ex-MORA beam size of 1.68\,arcsec$\times$1.44\,arcsec, this leads to an underestimate by a factor of 0.97, 0.91, and 0.71 for a source intrinsic FWHM of 0.25, 0.5, and 1\,arcsec, respectively. This effect is thus non negligible even if galaxies are 3 times more compact than the beam. The size of the sources can vary from a radius of $\sim$0.5\,kpc to $\sim$3\,kpc \citep{Gomez-Guijarro2022,Pozzi2024,Hodge2025}, corresponding to a FWHM from $\sim$0.1\,arcsec to $\sim$1\,arcsec for galaxies between z=1 to z=5. Depending on the source, the effect can range from negligible to strong.

The dotted line in Fig.\,\ref{fig:comp_mora} shows the expected relation between the N2CLS and Ex-MORA flux after applying all these corrections for a source FWHM of 0.5\,arcsec and a $\nu^{-2.95}$ reference spectrum. When all these effects are included, there is no longer a discrepancy between the two surveys. The comparison between single-dish and interferometric surveys is thus not straightforward. At similar depth, interferometric surveys will tend to miss extended and multi-component objects, while single-dish surveys will be biased towards groups of galaxies or random alignments. The significant number of N2CLS sources without an Ex-MORA counterpart suggests that these extended and composite sources are not rare. In addition, depending on the exact noise realization in each survey, some sources close to the detection limit may be detected by only one survey. This emphasizes the need for future combined analyses of surveys of the same area at different resolutions.

\section{Discussion on the relevance of 2\,mm continuum surveys to find high-z DSFGs}

\label{sect:discussion}

The colors and redshifts of the N2CLS 2\,mm sources agree with the predictions from the SIDES model (Sect.\,\ref{sect:colors}, \ref{sect:Nz}, and \ref{sect:col_vs_z}). In addition, the 2\,mm-only sources (only 8 out of 115 N2CLS 2\,mm sources for both GOODS-N and COSMOS) are either z$\lesssim$1 radio sources or are compatible with the expected spurious detection numbers. This tends to rule out the possibility of an exotic population of numerous galaxies detected only at 2\,mm at high $z$ or with very cold dust. A similar conclusion is reached by \citet{Ponthieu2025} based the measured N2CLS confusion noise, which is similar to the SIDES predictions. This is also consistent the results of the MORA and Ex-MORA surveys \citep{Casey2021,Zavala2021,Long2024}, which also found no excess of exotic very dusty sources. These new data combined with our model can also provide insights on the scientific performance (e.g., number of detections, coverage of the SFR-z plane) expected from 2\,mm surveys, and better understand the impact of the flux selection at this wavelength.

\subsection{Based on N2CLS observations}

\label{sect:disc_obs}

In this section, we discuss the merits of surveys at different wavelengths for detecting galaxies at various redshifts. Here, we consider a wavelength to be more efficient than another if it can detect a source with a higher S/N, or equivalently with less telescope time. Surveys can have different integration times and areas. It is thus important to find a common metric for all of them. In this discussion, we first focus on the case of comparing surveys in different bands (whether simultaneous or not) with the same area $\Omega_{\rm survey}$ and total time $t_{\rm survey}$. Since the sensitivity $\sigma$ in both bands scales as $t_{\rm survey}^{-1/2}$ at fixed area and as $\Omega_{\rm survey}^ {-1/2}$ at fixed total time, the sensitivity ratio between the two bands will be invariant as long as the two bands have the same area $\Omega_{\rm survey}$ and total time $t_{\rm survey}$. Sources with a color $S_A/S_B$ equal to the sensitivity ratio $\sigma_A / \sigma_B$ will have the same expected S/N in both bands ($S_A/\sigma_A = S_B/\sigma_B$). Sources with colors larger (lower, respectively) than the sensitivity ratio will be detected with a higher S/N in band A (band B, respectively). If most of the sources colors are above (below, respectively) the sensitivity ratio, the survey will be more efficient in band A (band B, respectively).

In Fig.\,\ref{fig:col_vs_z}, we added the N2CLS sensitivity ratio between 2\,mm and 1.2\,mm (left panels, dot-dash lines) to enable a comparison with the colors of N2CLS sources. We also computed the sensitivity ratio of ALMA using the observing tool\footnote{https://almascience.eso.org/proposing/observing-tool}. We estimated the expected sensitivity for 1.5$\times$1.5\,arcmin$^2$ continuum mosaic at 150\,GHz (2\,mm, band 4) and 250\,GHz (1.2\,mm, band 6) with similar total observing times (5.85\,h, 0.02\,mJy RMS at 2\,mm and 0.044\,mJy RMS at 1.2\,mm). The chosen size and the total time used in our computation are not important for estimating the sensitivity ratio, if we neglect the discontinuities in the total survey time generated by the need for new calibration cycles when going deeper. In contrast, this method takes into account that more pointings are necessary at higher frequency to cover the same area due to the smaller primary beam. For NIKA2, the 1.2\,mm is more efficient than the 2\,mm for most of the sources across the full redshift range covered by the N2CLS (z$\lesssim$6.5). For ALMA, the 1.2\,mm is consistently more efficient. A similar comparison can be made between the 2\,mm and 850\,$\mu$m bands. In the case of ALMA, the 850\,$\mu$m (353\,GHz, band ; 0.16\,mJy RMS on 1.5$\times$1.5\,arcmin$^2$ observed in 5.85\,h) observations are overall more efficient than those at 2\,mm (band 4, 150\,GHz). However, the weak trend of the color versus redshift (Sect.\,\ref{sect:col_vs_z}) suggests the possibility of a turnover in favor of the 2\,mm band at high redshift. We discuss this possibility based on the SIDES simulation in Sect.\,\ref{sect:disc_model}.


The comparison of surveys that use different observational strategies (e.g., total observing time or survey area) and different instruments is less direct. For example, for a fixed total observing time, a narrower and deeper survey may appear more efficient. Therefore, comparisons between different instruments (e.g., SCUBA2 and NIKA2) should be interpreted with caution. N2CLS 2\,mm could be considered more efficient than S2CLS 850\,$\mu$m \citep{Simpson2017} at all redshifts, although this comes with the caveat that S2CLS covers a larger area. In contrast, SUPER GOODS 850\,$\mu$m \citep{Cowie2017} is more efficient than the 2\,mm band of the N2CLS at z$\lesssim$4, but the trend is weak and the turnover redshift is highly uncertain. Overall, 1.2\,mm emerges as the most effective wavelength for conducting a census of dusty galaxies in deep blank cosmological surveys.

\begin{figure}
\centering
\includegraphics[width=9cm]{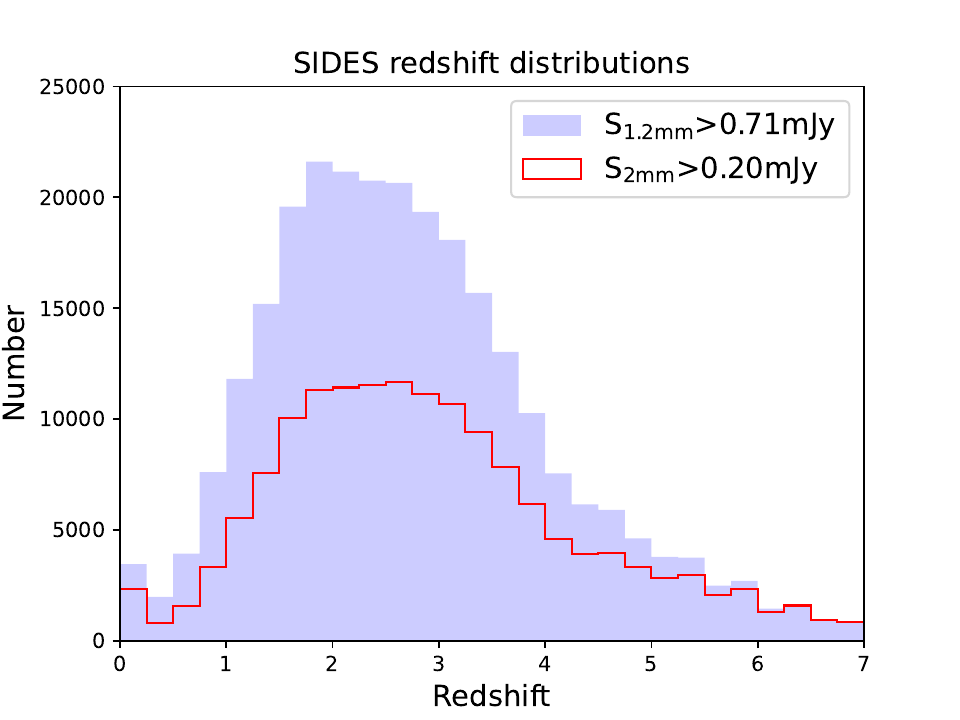}
\caption{\label{fig:Nz} Redshift distribution of the sources predicted by the 117\,deg$^2$ SIDES simulation for flux cuts at 1.2\,mm (blue filled histogram) and 2\,mm (red open histogram) similar to N2CLS in GOODS-N.}
\end{figure}

\begin{figure}
\centering
\includegraphics[width=9cm]{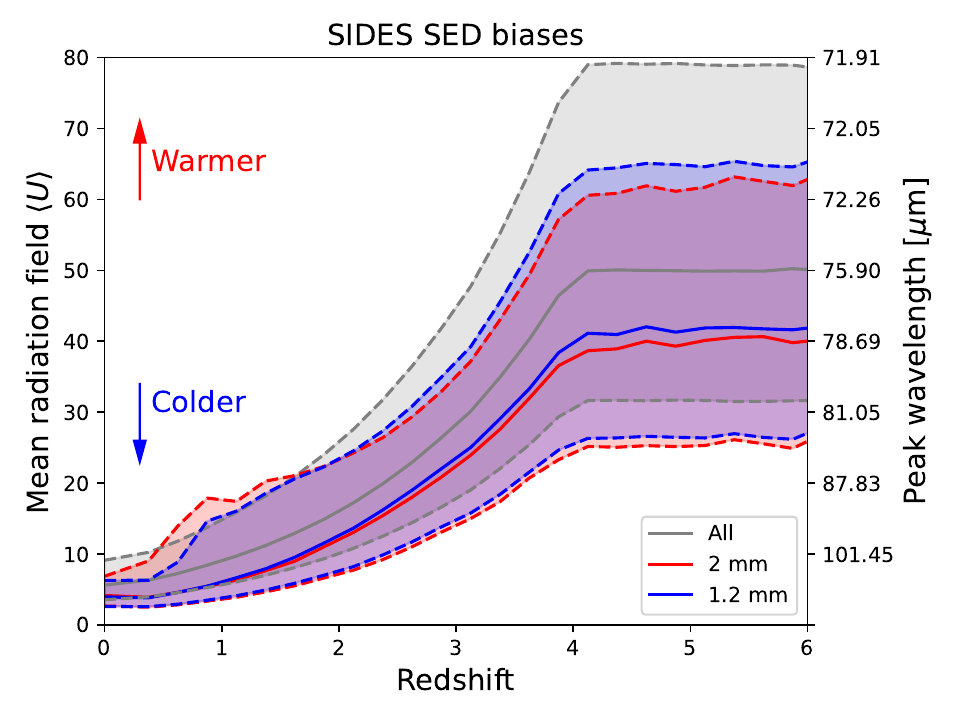}
\caption{\label{fig:Tdust_bias}  Mean interstellar radiation field $\langle U \rangle$ as a function of redshift, as predicted by SIDES, illustrating the bias toward colder SEDs introduced by millimeter flux selections. The mean radiation field is used in SIDES to parametrize the SEDs, and a higher $\langle U \rangle$ corresponds to a higher dust temperature. On the right y-axis, we also indicates the corresponding peak wavelength in S$_\nu$ units of the main-sequence SED templates. The solid lines are the mean evolutions as a function of redshift for the full star-forming sample (gray), and selections at 1.2\,mm (blue) and 2\,mm (red), while the shaded areas and the dashed lines correspond to the 16--84\% interval. }
\end{figure}

\subsection{Based on SIDES}

\label{sect:disc_model}

We can also use the SIDES simulation to estimate the most efficient band across the redshift range. The 2\,mm band becomes more efficient than the 1.2\,mm band only at z$\gtrsim$7.5 for NIKA2, and never in the case of ALMA. At these very high redshifts, we do not expect a significant population of heavily dust-obscured sources to be formed \citep[e.g.,][]{Michalowski2015,Burgarella2020}. Concerning the comparison between 2\,mm and 850\,$\mu$m, we confirm that the N2CLS 2\,mm band is more efficient than the SUPER GOODS 850\,$\mu$m data at z$\gtrsim$4, and than the S2COSMOS 850\,$\mu$m data at all redshifts. For ALMA, SIDES predicts that the turnover in favor of the 2\,mm occurs around z=5.5.

Since SIDES is able to reproduce the current 2\,mm observations (see also \citealt{Bing2023} for the number counts), we can use this model to better understand the impact of a 2\,mm selection. In Fig.\,\ref{fig:Nz}, we show the predicted redshift distribution of galaxies selected using similar flux limits as N2CLS in GOODS-N (see Sect.\,\ref{sect:Nz}). Contrary to previous observational redshift distributions, these distributions are not normalized, and indicate the total number of detections. This allows us to compare directly the number of expected detections in each redshift bin. The number of sources detected at 2\,mm is consistently lower than at 1.2\,mm across all bins covered by the simulation (0$<$z$<$7). As expected from the study of the color versus redshift, they start to be similar around z$\sim$7, where the number of detectable galaxies per redshift interval is a factor of 25 and 13 lower than at the peak around cosmic noon (z$\sim2.5$) at 1.2\,mm and 2\,mm, respectively. At cosmic noon, the number of galaxies detected at 2\,mm is approximately half of that detected at 1.2\,mm.

The higher mean redshift observed in the 2\,mm-selected sample is thus primarily due to the reduced number of galaxies detected around cosmic noon, compared to the 1.2\,mm selection, rather than an increased sensitivity to higher-redshift sources. \citet{Casey2021} argued that 2-mm surveys are a way to find the needles (z$>$4 objects) in the haystack (cosmic-noon sources). However, since most of the cosmic noon sources can be identified using ancillary data \citep{Berta2025}, a 1.2\,mm selection from which cosmic noon sources are removed is unlikely to produce more higher-z candidates to follow up than a simple 2\,mm flux selection. If we use only 2\,mm data, we lose information about the dust content of galaxies at cosmic noon \citep{Berta2025} without adding a significant number of new z$\gtrsim$4 targets. 

\citet{Cooper2022} promoted an alternative approach in which sources are selected at shorter wavelengths (850\,$\mu$m) from a single-dish survey and then followed up with ALMA at 2\,mm to reduce the uncertainties in sub-millimeter photometric redshifts from $\sigma_z / (1+z)$=0.3 to $\sigma_z / (1+z)$=0.2. However, this approach may be less efficient in the JWST era. For most sources, having the location of millimeter sources with a precision of $\lesssim1$\,arcsec is sufficient to find their JWST counterparts and obtain their photometric redshifts. The typical galaxy flux is a factor of $\sim$4 ($\sim$10, respectively) times higher at 1.2\,mm (850\,$\mu$m, respectively) than at 2\,mm, while ALMA single pointings are only a factor of 1.7 (2.2, respectively) times more sensitive at 2\,mm. We can thus follow up $\sim6$ ($\sim$20, respectively) times more sources with the same amount of telescope time at 1.2\,mm (850\,$\mu$m, respectively)\footnote{\rm At fixed integration time, the S/N ratio between 1.2\,mm and 2\,mm is $r = \frac{S_{\rm 1.2\,mm} \sigma_{\rm 2\,mm}}{S_{\rm 2\,mm} \sigma_{\rm 1.2\,mm}}$. Since S/N scales as $\sqrt{t}$, the time necessary to detect a source at a given S/N is thus shorter by a factor $r^2$ at 1.2\,mm.}. Therefore, it may thus be more appropriate to follow up the sources directly at shorter wavelengths (850\,$\mu$m or 1.2\,mm).

Finally, we used SIDES to understand how the selections at various wavelengths are biased towards or against different type of SEDs. The SIDES SEDs are parametrized using the mean radiation field $\langle U\rangle$. High values of this parameter correspond to high dust temperatures. In Fig.\,\ref{fig:Tdust_bias}, we show the redshift evolution of the mean $\langle U\rangle$ parameter together with the 16--84 percentile range corresponding to 1\,$\sigma$ in the Gaussian case. In SIDES, the $\langle U\rangle$ distribution varies with redshift but not with the stellar mass. We thus use the full simulated sample including low-mass galaxies\footnote{In SIDES, passive galaxies have a zero flux in both NIKA2 bands and the color is thus not defined. These objects are thus excluded from our analysis.} to provide our unbiased reference (in grey in the figure). We compare this sample with our flux selections at 1.2\,mm (blue) and 2\,mm (red) using the N2CLS GOODS-N flux cut. The mean values are biased towards lower $\langle U\rangle$ values (colder dust temperature) by 1\,$\sigma$ at z$<$1 and $\sim$0.5\,$\sigma$ at z$\gtrsim$4. The 2\,mm selection is slightly more biased toward high-redshift sources, by approximately 10\%.
This is not surprising, since colder dust SEDs emit more in the millimeter at fixed total dust luminosity. 

\begin{figure}
\centering
\includegraphics[width=9cm]{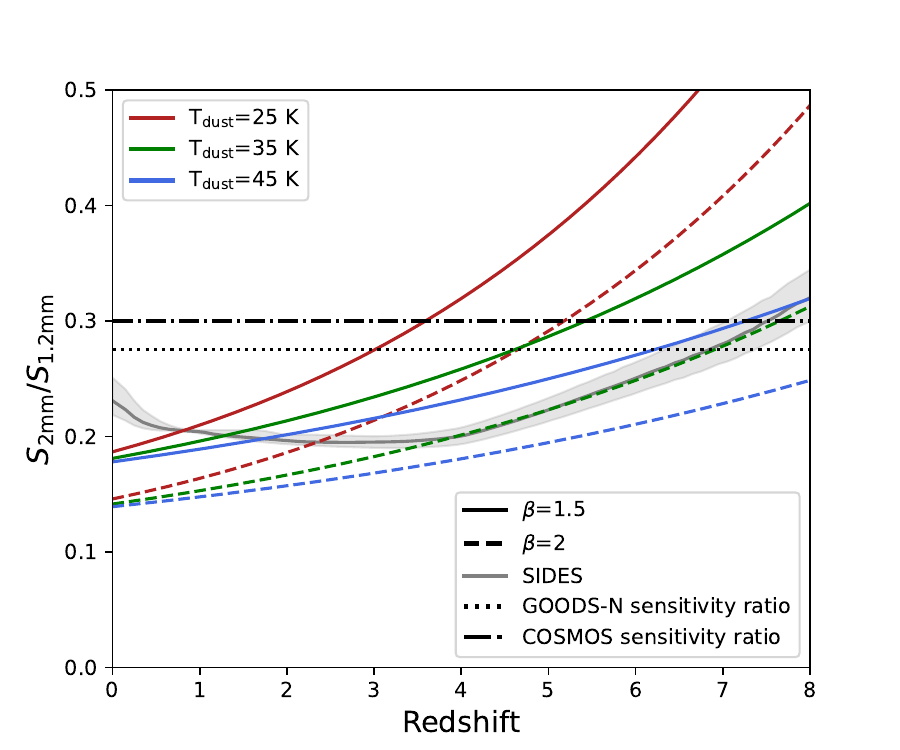}
\caption{\label{fig:colors_Tdust} Relation between the N2CLS color and the redshift for various SED models. The grey solid line and the associated contours represent the median color of SIDES galaxies and the 16--84\,\% dispersion. The solid and dashed colored lines (red for 25\,K, green for 35\,K, and blue for 45\,K) correspond to modified black-bodies with $\beta$=1.5 and $\beta$=2, respectively. The horizontal lines show the sensitivity ratio in COSMOS (dot-dash line) and GOODS-N (dotted line).}
\end{figure}

Our modeling analysis is based on the SIDES model, which does not contain a numerous population of dusty galaxies with very cold dust or at very high redshift. In Fig.\,\ref{fig:colors_Tdust}, we show the relation between color and redshift for both SIDES and several modified black-body SED models. For SIDES, the 1.2\,mm is systematically more sensitive than the 2\,mm up to z$\sim$7. We could wonder if some very cold sources missing in SIDES could be detected at 2\,mm but not at 1.2\,mm. To explore this hypothesis, we computed the NIKA2 color expected from modified black-bodies ($\nu^\beta \, B_\nu(T)$) with various dust temperatures T$_{\rm dust}$ and emissivity indices $\beta$. Galaxies with T$_{\rm dust}$=25\,K and $\beta$=1.5 at z$\gtrsim$3.5 are easier to detect at 2\,mm. Galaxies with T$_{\rm dust}$=35\,K and $\beta$=1.5 or T$_{\rm dust}$=25\,K and $\beta$=2 are detected first at 2\,mm for z$\gtrsim$5. For other SEDs with a higher T$_{\rm dust}$ or $\beta$, the 1.2\,mm is more efficient to detect them at least up to z$\sim$7. If dusty populations at z$\gtrsim$4 would be mainly objects with very cold dust or a low $\beta$, we would have found large populations of galaxies detected at 2\,mm or with high S$_{\rm 2\,mm}$/S$_{\rm 1.2\,mm}$ colors. Their absence in N2CLS is thus compatible with warmer SEDs, in line with other studies on the evolution the dust temperature with redshift based on small targeted ALMA samples \citep[e.g.,][]{Faisst2020b,Sommovigo2022} or stacking \citep[e.g.,][]{Bethermin2015,Schreiber2018,Viero2022}, which found an increasing T$_{\rm dust}$ with redshift.

\subsection{What is the most efficient approach for millimeter high-z surveys?}

The 2\,mm surveys as GISMO \citep{Staguhn2014,Magnelli2019}, N2CLS \citep{Bing2023,Ponthieu2025}, and (Ex-)MORA \citep{Casey2021,Long2024} have opened the 2\,mm window for deep and wide high-z galaxy surveys.  Unfortunately, these surveys did not reveal any new exotic population. However, they were very important to clarify which dusty sources could be found at z$\gtrsim$4. With all these data in hands, we can now conclude that the optimal wavelength range to probe the z$\gtrsim$4 dusty galaxy population is around 1.2\,mm, and future surveys should target it in priority. Coverage of the same sky area at 850\,$\mu$m and 2\,mm can help identify some high-redshift candidates; however, this identification remains uncertain due to large color measurement errors caused by limited signal-to-noise, the map making, and the source extraction. 

Having both interferometric and single-dish surveys provides an opportunity to better understand the impact of the angular resolution on the measurements (see Sect.\,\ref{sect:MORA}), and figure out the best observational strategy for future programs. Since high-z dusty sources have a low surface density with $\ll$1 source per ALMA or NOEMA pointing, probing a large volume could be more efficient using a single-dish mapping and a follow up of the detections with an interferometer to unambiguously identify their counterparts seen by, e.g., JWST.

We use Ex-MORA and N2CLS as examples to evaluate the merit of a direct interferometric survey against the single-dish approach followed up by interferometric observations. The number of detections per hour is comparable for (Ex-)MORA (37 detections in $\sim$44\,h, combining MORA and Ex-MORA) and N2CLS in COSMOS (90 detections in 195\,h). However, the carbon footprint (see Appendix\,\ref{sect:footprints}) and economical cost per hour of observation are dramatically lower for N2CLS at the IRAM 30\,m telescope ($\sim$2\,k€/h, private communication, $\sim$0.1\,tCO$_2$e/h) than (Ex-)MORA at ALMA ($\sim$60\,k€/h, $\sim$5\,tCO$_2$e/h). The follow up of all robust N2CLS sources in COSMOS would require only 90 ALMA or NOEMA pointings, while (Ex-)MORA is based on 4851 individual ALMA pointings. These follow-up observations are thus shorter by orders of magnitudes than a blind interferometric survey. The sum of the cost of a blind single-dish survey and its interferometric follow-up is much lower both on the ecological and economical point of view. This should not be viewed as a shortcoming of (Ex-)MORA, since its final results were published before N2CLS. Rather, this highlights the importance of carefully considering survey strategies when designing future programs. Our discussion applies only to the case of deep surveys. Since a wide range of scientific research can only be conducted with ALMA,  efficient use of its observing time is crucial for the scientific community.

In the future, the performance of ALMA will improve (e.g., larger bandpass increasing the continuum sensitivity, potential future multi-beam array). Meanwhile, new generations of single-dish instruments will emerge such as the upgraded version of NIKA2 or TolTEC \citep{Wilson2020}, as well as future single-dish telescopes (e.g., AtLAST, \citealt{Mroczkowski2025}) will emerge. Therefore, it will therefore be essential to regularly reassess the optimal survey strategy in light of the evolving instrumental landscape and the performance of both interferometric and single-dish facilities.

\section{Conclusion}

\label{sect:conclusion}

The N2CLS survey produced an unprecedented sample of 115 robust source detections selected at 2\,mm (25 in GOODS-N and 90 in COSMOS). This enabled us to constrain their statistical properties, test our models, and gain deeper insights into the types of sources selected by such surveys. Our main results are the following:
\begin{itemize}
\item We measured the mean S$_{\rm 2\,mm}$/S$_{\rm 1.2\,mm}$ color of sources detected at both 2\,mm and 1.2\,mm in the COSMOS field, and found 0.222$\pm$0.008. Our E2E simulations based on the SIDES model and the N2CLS data themselves show that the full observational process produces biases of $\sim$10\,\%. We also obtained a scatter of 0.070$\pm$0.010 from the observed fluxes. Our simulation shows that the observational process can be responsible for most of this scatter and the intrinsic value could be significantly lower.
\item We studied the redshift distribution of the 2\,mm sources in both fields. In GOODS-N, we find a mean redshift of 3.6$\pm$0.3, which is marginally higher than the prediction from SIDES (2.9$\pm$0.2). This high value in GOODS-N is driven by the N2CLS overdensity at z$\sim$5.2 \citep{Lagache2025}. In the COSMOS field, we find a mean redshift of 3.2$\pm$0.2, which is identical to that obtained in SIDES.
\item We found that the observed S$_{\rm 2\,mm}$/S$_{\rm 1.2\,mm}$ and S$_{\rm 2\,mm}$/S$_{\rm 850}$ colors have a large scatter and barely evolves from z=0 to z=6. This suggests that this color cannot be used as a reliable way to select high-z dusty galaxies.
\item In COSMOS, there are 8 sources detected at 2\,mm without counterpart at 1.2\,mm. Two of these sources are radio galaxies at z$<$1. The six other sources are compatible with the expected number of spurious detections. In GOODS-N, only one 2\,mm source has no 1.2\,mm counterpart, but it is firmly identified as a z$\sim$2 galaxy thanks to NOEMA follow-up observations.
\item We compared the N2CLS 2\,mm flux measurements with the Ex-MORA ALMA survey. We observe a NIKA2 flux excess relative to ALMA, not only for sources with multiple components but also for those with a single component, with an average excess factor of 1.3. This is explained by a combination of multiple factors such as the different bandpass of the two surveys, the source blending, and the peak flux photometry used by Ex-MORA.
\item We discussed the relevance of 2\,mm surveys using both N2CLS and SIDES. We found that the slightly higher mean redshift of 2\,mm surveys compared to 1.2\,mm is mainly caused by a massive loss of cosmic noon sources, without any significant gain even at z$\gtrsim$5. The 2\,mm sample is also slightly more biased towards colder dust temperatures. We finally show that 2\,mm single-dish surveys are more efficient than ALMA surveys.
\end{itemize}

Although the deep 2\,mm N2CLS data could have revealed exotic populations of dusty galaxies with very cold dust temperatures or at extremely high redshifts, we found no evidence for such sources. Ruling this out is important for optimizing the wavelength selection of future surveys targeting dusty star-forming galaxies. Our results show that 1.2\,mm surveys are more efficient for building large samples from cosmic noon (z$\sim$2) to the reionization era (z$\sim$8).



\begin{acknowledgements}
We thank Arianna Long for providing various Ex-MORA products allowing us to compare our results. We thank Jürgen Knödlseder for providing us the carbon footprint estimates of ALMA and the 30\,m telescopes.

We acknowledge financial support from the Programme National de Cosmologie and Galaxies (PNCG) funded by CNRS/INSU-IN2P3-INP, CEA and CNES, France, and from the European Research Council (ERC) under the European Union's Horizon 2020 research and innovation programme (project CONCERTO, grant agreement No 788212).

This work is based on observations carried out under project numbers 192-16 with the IRAM 30-m telescope, and projects W21CV, W23CX, and S24CF with NOEMA. IRAM is supported by INSU/CNRS (France), MPG (Germany) and IGN (Spain). 

We would like to thank the IRAM staff for their support during the NIKA and NIKA2 campaigns. The NIKA2 dilution cryostat has been designed and built at the Institut N\'eel. In particular, we acknowledge the crucial contribution of the Cryogenics Group, and in particular Gregory Garde, Henri Rodenas, Jean Paul Leggeri, Philippe Camus. This work has been partially funded by the Foundation Nanoscience Grenoble and the LabEx FOCUS ANR-11-LABX-0013. This work is supported by the French National Research Agency under the contracts "MKIDS", "NIKA" and ANR-15-CE31-0017 and in the framework of the "Investissements d'avenir program (ANR-15-IDEX-02). This work has benefited from the support of the European Research Council Advanced Grant ORISTARS under the European Union's Seventh Framework Programme (Grant Agreement no. 291294). F.R. acknowledges financial supports provided by NASA through SAO Award Number SV2-82023 issued by the Chandra X-Ray Observatory Center, which is operated by the Smithsonian Astrophysical Observatory for and on behalf of NASA under contract NAS8-03060. 
The NIKA2 data were processed using the Pointing and Imaging In Continuum software \citep[PIIC,][]{Zylka2013,piic2024}, developed by Robert Zylka at the Institut de Radioastronomie Millimetrique (IRAM) and distributed by IRAM via the GILDAS pages. PIIC is the extension of the MOPSIC data reduction software to the case of NIKA2 data.
The National Radio Astronomy Observatory is a facility of the National Science Foundation operated under cooperative agreement by Associated Universities, Inc.
This work made use of Astropy:\footnote{http://www.astropy.org} a community-developed core Python package and an ecosystem of tools and resources for astronomy \citep{astropy:2013, astropy:2018, astropy:2022}. 

R.A. was supported by the French government through the France 2030 investment plan managed by the National Research Agency (ANR), as part of the Initiative of Excellence of Université Côte d’Azur under reference number ANR-15-IDEX-01.

\end{acknowledgements}

\bibliographystyle{aa}

\bibliography{biblio}

\begin{appendix}

\section{Estimate of the carbon footprint of ALMA and the IRAM 30\,m telescope}

\label{sect:footprints}

\citet{Knodlseder2024} estimated the carbon footprint of worldwide astronomical research facilities. They provided us their best estimate of the yearly carbon footprint of ALMA and the IRAM 30\,m telescope on which the NIKA2 camera is installed. For ALMA, they estimated 256\,000\,tCO$_2$e for the construction and 21\,000\,tCO$_2$e/yr for the operations. For the 30\,m, the construction produced 4700\,tCO$_2$e and the operations 300\,tCO$_2$e. If we distribute the construction impact over an assumed 50\,yr of operation, we obtain 26\,000\,tCO$_2$e/yr for ALMA and 400\,tCO$_2$e/yr. for the 30\,m. There is thus a ratio of 65 between the two carbon footprint. Both observatories announce $\sim$4000-5000\,h of useful science data per year, although the 30\,m tends to observe at lower frequency. Even when trying to correct for the better weather at ALMA for 2\,mm observations, it is rather safe to assume that the hourly carbon footprint ratio between the two facilites is at least 1.5 orders of magnitude higher.

\section{Tables}

The N2CLS robust (95\,\% purity) 2\,mm catalogs are presented in Table\,\ref{tab:cat_goodsn} for GOODS-N and Table\,\ref{tab:cat_cosmos} and \ref{tab:cat_cosmos2} for COSMOS. All the data used in this paper (1.2 and 2\,mm fluxes from N2CLS, 850\,$\mu$m fluxes from SCUBA2, redshifts, and simulations) can be accessed at \url{https://data.lam.fr/n2cls/}.

\begin{table*}
\centering
\caption{\label{tab:cat_goodsn} High-quality (95\,\% purity) 2\,mm N2CLS catalog in the GOODS-N field. The first and second columns are the N2CLS short names and the IAU names, respectively. The R.A.$_{\rm 2\,mm}$ and Dec$_{\rm 2\,mm}$ columns are the coordinates of the sources measured in the 2\,mm maps. The four last columns (S/N$_{\rm 2\,mm}$, S$_{\rm 2\,mm}$, S/N$_{\rm 1.2\,mm}$ et S$_{\rm 1.2\,mm}$) are the S/N at 2\,mm, the N2CLS flux at 2\,mm, the S/N at 1.2\,mm, and the N2CLS flux at 1.2\,mm, respectively. For the source without 1.2\,mm counterpart, we provide a 5\,$\sigma$ upper limit on the flux. The redshifts are provided in \citet{Berta2025}.}
\begin{tabular}{llllllll}
\hline
\hline
N2CLS name & IAU name & R.A.$_{\rm 2\,mm}$ & Dec$_{\rm 2\,mm}$ & S/N$_{\rm 2\,mm}$ & S$_{\rm 2\,mm}$ & S/N$_{\rm 1.2\,mm}$ & S$_{\rm 1.2\,mm}$ \\
& &  deg & deg & & mJy & & mJy \\
\hline
N2GN\_1\_01 & N2GN J123633+621408 & 189.1400 & 62.2361 & 35.7 & 1.22$\substack{0.13 \\ -0.12}$ & 40.0 & 5.00$\substack{0.48 \\ -0.48}$\\
N2GN\_1\_02 & N2GN J123730+621259 & 189.3786 & 62.2160 & 16.6 & 0.86$\substack{0.15 \\ -0.13}$ & 25.6 & 4.77$\substack{0.43 \\ -0.58}$\\
N2GN\_1\_03 & N2GN J123707+621408 & 189.2802 & 62.2355 & 15.3 & 0.52$\substack{0.07 \\ -0.09}$ & 23.2 & 2.91$\substack{0.30 \\ -0.22}$\\
N2GN\_1\_04 & N2GN J123711+622211 & 189.2991 & 62.3700 & 18.8 & 1.85$\substack{0.29 \\ -0.23}$ & 21.4 & 8.06$\substack{0.68 \\ -0.91}$\\
N2GN\_1\_05 & N2GN J123711+621330 & 189.2988 & 62.2253 & 16.0 & 0.60$\substack{0.10 \\ -0.08}$ & 20.6 & 2.71$\substack{0.28 \\ -0.31}$\\
N2GN\_1\_06 & N2GN J123652+621226 & 189.2160 & 62.2072 & 13.3 & 0.46$\substack{0.09 \\ -0.08}$ & 19.1 & 2.36$\substack{0.23 \\ -0.29}$\\
N2GN\_1\_07 & N2GN J123645+621448 & 189.1923 & 62.2470 & 9.3 & 0.31$\substack{0.06 \\ -0.09}$ & 19.0 & 2.29$\substack{0.24 \\ -0.22}$\\
N2GN\_1\_08 & N2GN J123631+621714 & 189.1333 & 62.2875 & 11.3 & 0.56$\substack{0.12 \\ -0.11}$ & 18.0 & 3.38$\substack{0.42 \\ -0.43}$\\
N2GN\_1\_09 & N2GN J123627+621217 & 189.1139 & 62.2046 & 10.6 & 0.36$\substack{0.08 \\ -0.08}$ & 18.0 & 2.20$\substack{0.31 \\ -0.25}$\\
N2GN\_1\_11 & N2GN J123713+621826 & 189.3081 & 62.3072 & 13.1 & 0.49$\substack{0.08 \\ -0.09}$ & 15.2 & 2.14$\substack{0.33 \\ -0.29}$\\
N2GN\_1\_12 & N2GN J123636+621155 & 189.1530 & 62.1985 & 8.3 & 0.27$\substack{0.10 \\ -0.07}$ & 15.1 & 1.79$\substack{0.25 \\ -0.29}$\\
N2GN\_1\_13 & N2GN J123658+621451 & 189.2440 & 62.2476 & 11.7 & 0.37$\substack{0.07 \\ -0.08}$ & 14.6 & 1.71$\substack{0.21 \\ -0.25}$\\
N2GN\_1\_14 & N2GN J123701+621146 & 189.2567 & 62.1963 & 6.7 & 0.28$\substack{0.10 \\ -0.09}$ & 13.3 & 1.88$\substack{0.25 \\ -0.27}$\\
N2GN\_1\_15 & N2GN J123618+621550 & 189.0762 & 62.2640 & 6.6 & 0.32$\substack{0.10 \\ -0.12}$ & 13.2 & 2.38$\substack{0.43 \\ -0.35}$\\
N2GN\_1\_16 & N2GN J123622+621615 & 189.0930 & 62.2721 & 4.7 & 0.23$\substack{0.13 \\ -0.12}$ & 12.5 & 2.33$\substack{0.37 \\ -0.35}$\\
N2GN\_1\_17 & N2GN J123738+621734 & 189.4097 & 62.2934 & 11.4 & 0.51$\substack{0.10 \\ -0.11}$ & 11.9 & 2.00$\substack{0.31 \\ -0.42}$\\
N2GN\_1\_18 & N2GN J123702+621425 & 189.2619 & 62.2413 & 6.7 & 0.21$\substack{0.08 \\ -0.09}$ & 11.6 & 1.36$\substack{0.27 \\ -0.21}$\\
N2GN\_1\_23 & N2GN J123656+621207 & 189.2333 & 62.2023 & 4.7 & 0.17$\substack{0.11 \\ -0.09}$ & 9.5 & 1.23$\substack{0.31 \\ -0.29}$\\
N2GN\_1\_24 & N2GN J123719+621219 & 189.3289 & 62.2048 & 4.2 & 0.20$\substack{0.18 \\ -0.12}$ & 8.9 & 1.49$\substack{0.34 \\ -0.37}$\\
N2GN\_1\_25 & N2GN J123712+621212 & 189.3037 & 62.2026 & 4.6 & 0.20$\substack{0.13 \\ -0.11}$ & 8.7 & 1.32$\substack{0.34 \\ -0.26}$\\
N2GN\_1\_28 & N2GN J123728+621920 & 189.3638 & 62.3227 & 5.8 & 0.27$\substack{0.13 \\ -0.13}$ & 7.9 & 1.38$\substack{0.39 \\ -0.36}$\\
N2GN\_1\_34 & N2GN J123644+621938 & 189.1864 & 62.3275 & 5.6 & 0.36$\substack{0.19 \\ -0.14}$ & 6.9 & 1.63$\substack{0.61 \\ -0.50}$\\
N2GN\_1\_36 & N2GN J123658+620930 & 189.2443 & 62.1585 & 6.4 & 0.36$\substack{0.14 \\ -0.14}$ & 6.0 & 1.19$\substack{0.43 \\ -0.38}$\\
N2GN\_2\_13 & N2GN J123720+621128 & 189.3374 & 62.1913 & 7.4 & 0.43$\substack{0.11 \\ -0.14}$ &  -- & $<$2\\
N2GN\_2\_20 & N2GN J123608+621251 & 189.0370 & 62.2142 & 5.1 & 0.24$\substack{0.13 \\ -0.12}$ & 3.3 & 0.52$\substack{0.29 \\ -0.28}$\\
\hline
\end{tabular}
\end{table*}

\begin{table*}
\centering
\caption{\label{tab:cat_cosmos} High-quality (95\,\% purity) 2\,mm N2CLS catalog in the GOODS-N field. The first and second columns are the N2CLS short names and the IAU names, respectively. The R.A.$_{\rm 2\,mm}$ and Dec$_{\rm 2\,mm}$ columns are the coordinates of the sources measured in the 2\,mm maps. The four last columns (S/N$_{\rm 2\,mm}$, S$_{\rm 2\,mm}$, S/N$_{\rm 1.2\,mm}$ et S$_{\rm 1.2\,mm}$) are the S/N at 2\,mm, the N2CLS flux at 2\,mm, the S/N at 1.2\,mm, and the N2CLS flux at 1.2\,mm, respectively. For the source without 1.2\,mm counterpart, we provide a 5\,$\sigma$ upper limit on the flux. The redshifts are provided in \citet{Berta2025}.}
\begin{tabular}{llllllll}
\hline
\hline
N2CLS name & IAU name & R.A.$_{\rm 2\,mm}$ & Dec$_{\rm 2\,mm}$ & S/N$_{\rm 2\,mm}$ & S$_{\rm 2\,mm}$ & S/N$_{\rm 1.2\,mm}$ & S$_{\rm 1.2\,mm}$ \\
& &  deg & deg & & mJy & & mJy \\
\hline
N2CO\_1\_1 & N2CO J100008+022612 & 150.0332 & 2.4369 & 22.0 & 2.07$\substack{0.13 \\ -0.12}$ & 28.1 & 9.20$\substack{0.52 \\ -0.64}$ \\
N2CO\_1\_2 & N2CO J100015+021549 & 150.0647 & 2.2638 & 17.1 & 1.62$\substack{0.13 \\ -0.13}$ & 22.6 & 7.18$\substack{0.48 \\ -0.58}$ \\
N2CO\_1\_3 & N2CO J095942+022937 & 149.9283 & 2.4940 & 14.3 & 1.38$\substack{0.16 \\ -0.16}$ & 22.4 & 7.71$\substack{0.50 \\ -0.62}$ \\
N2CO\_1\_4 & N2CO J100057+022014 & 150.2379 & 2.3372 & 16.0 & 1.46$\substack{0.15 \\ -0.13}$ & 21.8 & 6.84$\substack{0.46 \\ -0.45}$ \\
N2CO\_1\_5 & N2CO J095957+022731 & 149.9886 & 2.4588 & 16.1 & 1.55$\substack{0.15 \\ -0.14}$ & 20.4 & 6.87$\substack{0.48 \\ -0.44}$ \\
N2CO\_1\_6 & N2CO J100122+022005 & 150.3459 & 2.3346 & 11.9 & 1.49$\substack{0.22 \\ -0.20}$ & 17.7 & 7.18$\substack{0.63 \\ -0.65}$ \\
N2CO\_1\_7 & N2CO J100019+023204 & 150.0818 & 2.5349 & 9.4 & 0.87$\substack{0.17 \\ -0.14}$ & 17.0 & 5.49$\substack{0.49 \\ -0.49}$ \\
N2CO\_1\_8\tablefootmark{a} & N2CO J095959+023442 & 149.9966 & 2.5797 & 12.0 & 1.42$\substack{0.21 \\ -0.20}$ & 15.2 & 10.24$\substack{0.88 \\ -0.86}$ \\
N2CO\_1\_9 & N2CO J100028+023204 & 150.1195 & 2.5347 & 9.2 & 0.88$\substack{0.17 \\ -0.16}$ & 14.2 & 4.66$\substack{0.54 \\ -0.52}$ \\
N2CO\_1\_10 & N2CO J100031+021241 & 150.1326 & 2.2115 & 8.7 & 0.81$\substack{0.16 \\ -0.15}$ & 14.0 & 4.36$\substack{0.52 \\ -0.50}$ \\
N2CO\_1\_11 & N2CO J100043+020519 & 150.1797 & 2.0890 & 5.3 & 0.50$\substack{0.16 \\ -0.14}$ & 13.7 & 4.39$\substack{0.53 \\ -0.49}$ \\
N2CO\_1\_12 & N2CO J100121+023129 & 150.3404 & 2.5253 & 5.6 & 0.68$\substack{0.21 \\ -0.19}$ & 12.9 & 5.23$\substack{0.63 \\ -0.61}$ \\
N2CO\_1\_13 & N2CO J100034+020302 & 150.1429 & 2.0509 & 6.4 & 0.61$\substack{0.17 \\ -0.15}$ & 12.7 & 4.10$\substack{0.51 \\ -0.50}$ \\
N2CO\_1\_14 & N2CO J095959+020633 & 149.9995 & 2.1096 & 7.2 & 0.67$\substack{0.16 \\ -0.14}$ & 12.6 & 3.93$\substack{0.50 \\ -0.46}$ \\
N2CO\_1\_16 & N2CO J100025+022606 & 150.1048 & 2.4345 & 7.4 & 0.70$\substack{0.16 \\ -0.15}$ & 12.3 & 3.93$\substack{0.52 \\ -0.51}$ \\
N2CO\_1\_17 & N2CO J100023+021750 & 150.1004 & 2.2977 & 7.3 & 0.69$\substack{0.16 \\ -0.15}$ & 12.2 & 3.94$\substack{0.52 \\ -0.51}$ \\
N2CO\_1\_18 & N2CO J100033+022600 & 150.1386 & 2.4335 & 7.2 & 0.67$\substack{0.16 \\ -0.14}$ & 11.9 & 3.88$\substack{0.53 \\ -0.51}$ \\
N2CO\_1\_19 & N2CO J100004+023046 & 150.0205 & 2.5127 & 6.0 & 0.56$\substack{0.16 \\ -0.15}$ & 11.6 & 3.70$\substack{0.50 \\ -0.48}$ \\
N2CO\_1\_20 & N2CO J100010+021335 & 150.0419 & 2.2266 & 7.8 & 0.72$\substack{0.16 \\ -0.15}$ & 11.5 & 3.61$\substack{0.49 \\ -0.48}$ \\
N2CO\_1\_21 & N2CO J100033+020850 & 150.1405 & 2.1474 & 8.8 & 0.83$\substack{0.16 \\ -0.16}$ & 11.5 & 3.84$\substack{0.52 \\ -0.50}$ \\
N2CO\_1\_22 & N2CO J100038+020823 & 150.1584 & 2.1394 & 6.5 & 0.60$\substack{0.17 \\ -0.15}$ & 11.5 & 3.65$\substack{0.49 \\ -0.49}$ \\
N2CO\_1\_23 & N2CO J100021+020041 & 150.0887 & 2.0115 & 10.8 & 1.26$\substack{0.22 \\ -0.20}$ & 11.0 & 4.47$\substack{0.63 \\ -0.59}$ \\
N2CO\_1\_24 & N2CO J100029+020527 & 150.1205 & 2.0901 & 7.9 & 0.75$\substack{0.17 \\ -0.16}$ & 11.0 & 3.54$\substack{0.49 \\ -0.47}$ \\
N2CO\_1\_25 & N2CO J100013+023428 & 150.0556 & 2.5736 & 9.6 & 1.14$\substack{0.21 \\ -0.19}$ & 10.8 & 4.34$\substack{0.60 \\ -0.59}$ \\
N2CO\_1\_26 & N2CO J100025+021847 & 150.1039 & 2.3127 & 5.2 & 0.47$\substack{0.19 \\ -0.14}$ & 10.8 & 3.46$\substack{0.53 \\ -0.53}$ \\
N2CO\_1\_28 & N2CO J100104+022858 & 150.2712 & 2.4822 & 5.9 & 0.57$\substack{0.18 \\ -0.16}$ & 10.6 & 3.52$\substack{0.54 \\ -0.54}$ \\
N2CO\_1\_29 & N2CO J095931+023044 & 149.8826 & 2.5129 & 10.1 & 1.40$\substack{0.26 \\ -0.23}$ & 10.6 & 4.81$\substack{0.80 \\ -0.77}$ \\
N2CO\_1\_30 & N2CO J100023+022155 & 150.0984 & 2.3647 & 6.4 & 0.61$\substack{0.17 \\ -0.15}$ & 10.4 & 3.42$\substack{0.55 \\ -0.51}$ \\
N2CO\_1\_31 & N2CO J100020+023520 & 150.0855 & 2.5894 & 6.2 & 0.83$\substack{0.24 \\ -0.20}$ & 10.2 & 4.71$\substack{0.73 \\ -0.71}$ \\
N2CO\_1\_32 & N2CO J100025+020313 & 150.1061 & 2.0544 & 5.8 & 0.56$\substack{0.17 \\ -0.15}$ & 10.2 & 3.31$\substack{0.53 \\ -0.50}$ \\
N2CO\_1\_33 & N2CO J100114+022705 & 150.3109 & 2.4502 & 6.7 & 0.64$\substack{0.18 \\ -0.15}$ & 10.1 & 3.30$\substack{0.52 \\ -0.52}$ \\
N2CO\_1\_34 & N2CO J100049+022258 & 150.2079 & 2.3834 & 7.2 & 0.68$\substack{0.16 \\ -0.15}$ & 10.0 & 3.37$\substack{0.53 \\ -0.53}$ \\
N2CO\_1\_35 & N2CO J100059+021716 & 150.2462 & 2.2874 & 5.8 & 0.55$\substack{0.17 \\ -0.15}$ & 9.8 & 3.14$\substack{0.53 \\ -0.48}$ \\
N2CO\_1\_36 & N2CO J100015+022445 & 150.0660 & 2.4127 & 5.2 & 0.50$\substack{0.16 \\ -0.14}$ & 9.8 & 3.18$\substack{0.50 \\ -0.50}$ \\
N2CO\_1\_39 & N2CO J100007+021148 & 150.0321 & 2.1970 & 6.6 & 0.64$\substack{0.17 \\ -0.15}$ & 9.5 & 3.15$\substack{0.50 \\ -0.51}$ \\
N2CO\_1\_40 & N2CO J100008+021306 & 150.0362 & 2.2191 & 5.3 & 0.52$\substack{0.16 \\ -0.15}$ & 9.5 & 3.09$\substack{0.50 \\ -0.49}$ \\
N2CO\_1\_41\tablefootmark{a} & N2CO J100027+023137 & 150.1118 & 2.5252 & 9.2 & 0.86$\substack{0.16 \\ -0.15}$ & 9.2 & 3.41$\substack{0.59 \\ -0.54}$ \\
N2CO\_1\_42 & N2CO J100111+022841 & 150.2982 & 2.4780 & 6.9 & 0.68$\substack{0.18 \\ -0.16}$ & 9.2 & 3.16$\substack{0.58 \\ -0.51}$ \\
N2CO\_1\_45 & N2CO J100025+020051 & 150.1059 & 2.0143 & 7.7 & 0.90$\substack{0.20 \\ -0.19}$ & 9.2 & 3.74$\substack{0.68 \\ -0.62}$ \\
N2CO\_1\_46 & N2CO J100054+023435 & 150.2275 & 2.5768 & 6.6 & 0.81$\substack{0.21 \\ -0.19}$ & 9.1 & 3.79$\substack{0.67 \\ -0.62}$ \\
N2CO\_1\_47 & N2CO J100102+022236 & 150.2591 & 2.3771 & 6.5 & 0.61$\substack{0.17 \\ -0.15}$ & 9.1 & 2.96$\substack{0.52 \\ -0.48}$ \\
N2CO\_1\_49 & N2CO J100026+021528 & 150.1098 & 2.2576 & 6.7 & 0.63$\substack{0.18 \\ -0.15}$ & 9.0 & 3.01$\substack{0.54 \\ -0.50}$ \\
N2CO\_1\_50 & N2CO J100012+020124 & 150.0524 & 2.0243 & 8.2 & 0.87$\substack{0.18 \\ -0.17}$ & 9.0 & 3.38$\substack{0.62 \\ -0.56}$ \\
N2CO\_1\_51 & N2CO J100024+022005 & 150.1015 & 2.3352 & 5.1 & 0.47$\substack{0.18 \\ -0.14}$ & 8.9 & 2.90$\substack{0.54 \\ -0.47}$ \\
N2CO\_1\_53 & N2CO J100104+020203 & 150.2685 & 2.0338 & 5.1 & 0.50$\substack{0.19 \\ -0.15}$ & 8.6 & 3.00$\substack{0.55 \\ -0.51}$ \\
N2CO\_1\_54 & N2CO J095952+022139 & 149.9690 & 2.3599 & 4.8 & 0.44$\substack{0.18 \\ -0.14}$ & 8.6 & 2.73$\substack{0.52 \\ -0.47}$ \\
N2CO\_1\_55 & N2CO J100047+021016 & 150.1966 & 2.1708 & 5.6 & 0.53$\substack{0.16 \\ -0.14}$ & 8.4 & 2.82$\substack{0.53 \\ -0.48}$ \\
N2CO\_1\_58 & N2CO J100047+020938 & 150.1967 & 2.1604 & 8.5 & 0.77$\substack{0.15 \\ -0.15}$ & 8.4 & 2.66$\substack{0.50 \\ -0.46}$ \\
N2CO\_1\_59 & N2CO J100105+022132 & 150.2730 & 2.3592 & 5.4 & 0.36$\substack{0.13 \\ -0.12}$ & 8.3 & 2.65$\substack{0.51 \\ -0.46}$ \\
\hline
\end{tabular}
\tablefoot{
\tablefoottext{a}{These 2\,mm sources have two 1.2\,mm counterparts. In our analysis, we use the sum of the two 1.2\,mm sources to compute their color. The uncertainties are combined quadratically.}}
\end{table*}

\begin{table*}
\centering
\caption{\label{tab:cat_cosmos2} Continuation of Table\,\ref{tab:cat_cosmos}.}
\begin{tabular}{llllllll}
\hline
\hline
N2CLS name & IAU name & R.A.$_{\rm 2\,mm}$ & Dec$_{\rm 2\,mm}$ & S/N$_{\rm 2\,mm}$ & S$_{\rm 2\,mm}$ & S/N$_{\rm 1.2\,mm}$ & S$_{\rm 1.2\,mm}$ \\
& &  deg & deg & & mJy & & mJy \\
\hline
N2CO\_1\_60 & N2CO J100010+022223 & 150.0424 & 2.3737 & 5.3 & 0.51$\substack{0.16 \\ -0.14}$ & 8.3 & 2.68$\substack{0.50 \\ -0.46}$ \\
N2CO\_1\_62 & N2CO J100012+021211 & 150.0544 & 2.2029 & 5.9 & 0.55$\substack{0.17 \\ -0.15}$ & 8.3 & 2.71$\substack{0.52 \\ -0.46}$ \\
N2CO\_1\_63 & N2CO J095950+022827 & 149.9610 & 2.4743 & 4.9 & 0.44$\substack{0.18 \\ -0.14}$ & 8.3 & 2.62$\substack{0.51 \\ -0.44}$ \\
N2CO\_1\_65 & N2CO J100112+020852 & 150.3008 & 2.1477 & 8.5 & 0.86$\substack{0.16 \\ -0.17}$ & 8.2 & 2.85$\substack{0.55 \\ -0.49}$ \\
N2CO\_1\_67 & N2CO J095953+021853 & 149.9721 & 2.3147 & 7.8 & 0.72$\substack{0.16 \\ -0.15}$ & 8.0 & 2.58$\substack{0.54 \\ -0.51}$ \\
N2CO\_1\_71 & N2CO J100104+022634 & 150.2693 & 2.4427 & 4.7 & 0.43$\substack{0.17 \\ -0.14}$ & 7.8 & 2.58$\substack{0.54 \\ -0.49}$ \\
N2CO\_1\_72 & N2CO J100024+022947 & 150.1001 & 2.4962 & 7.0 & 0.65$\substack{0.15 \\ -0.14}$ & 7.7 & 2.53$\substack{0.54 \\ -0.49}$ \\
N2CO\_1\_74 & N2CO J100057+021309 & 150.2393 & 2.2193 & 6.7 & 0.63$\substack{0.17 \\ -0.15}$ & 7.7 & 2.50$\substack{0.53 \\ -0.48}$ \\
N2CO\_1\_76 & N2CO J100105+022151 & 150.2718 & 2.3631 & 5.4 & 0.34$\substack{0.14 \\ -0.11}$ & 7.7 & 2.41$\substack{0.51 \\ -0.47}$ \\
N2CO\_1\_78 & N2CO J100117+023218 & 150.3230 & 2.5378 & 6.0 & 0.64$\substack{0.19 \\ -0.17}$ & 7.6 & 2.73$\substack{0.59 \\ -0.52}$ \\
N2CO\_1\_79 & N2CO J100058+020138 & 150.2436 & 2.0285 & 6.1 & 0.62$\substack{0.17 \\ -0.15}$ & 7.5 & 2.72$\substack{0.57 \\ -0.53}$ \\
N2CO\_1\_85 & N2CO J100041+022547 & 150.1749 & 2.4291 & 4.6 & 0.42$\substack{0.17 \\ -0.14}$ & 7.4 & 2.37$\substack{0.53 \\ -0.46}$ \\
N2CO\_1\_87 & N2CO J100109+021727 & 150.2907 & 2.2897 & 4.8 & 0.44$\substack{0.17 \\ -0.14}$ & 7.3 & 2.35$\substack{0.52 \\ -0.46}$ \\
N2CO\_1\_88 & N2CO J100056+020842 & 150.2342 & 2.1450 & 5.2 & 0.49$\substack{0.18 \\ -0.15}$ & 7.3 & 2.37$\substack{0.52 \\ -0.47}$ \\
N2CO\_1\_93 & N2CO J100106+021532 & 150.2783 & 2.2587 & 5.2 & 0.50$\substack{0.19 \\ -0.15}$ & 7.2 & 2.45$\substack{0.54 \\ -0.48}$ \\
N2CO\_1\_97 & N2CO J100013+020902 & 150.0534 & 2.1495 & 5.9 & 0.57$\substack{0.18 \\ -0.15}$ & 7.1 & 2.39$\substack{0.52 \\ -0.48}$ \\
N2CO\_1\_99 & N2CO J100118+022352 & 150.3279 & 2.3980 & 4.8 & 0.51$\substack{0.20 \\ -0.16}$ & 7.0 & 2.56$\substack{0.55 \\ -0.52}$ \\
N2CO\_1\_101 & N2CO J100046+021309 & 150.1923 & 2.2197 & 7.9 & 0.75$\substack{0.17 \\ -0.16}$ & 6.9 & 2.25$\substack{0.57 \\ -0.51}$ \\
N2CO\_1\_108 & N2CO J095944+022109 & 149.9344 & 2.3530 & 7.0 & 0.68$\substack{0.18 \\ -0.16}$ & 6.6 & 2.23$\substack{0.57 \\ -0.51}$ \\
N2CO\_1\_111 & N2CO J100018+021241 & 150.0769 & 2.2107 & 4.8 & 0.45$\substack{0.18 \\ -0.14}$ & 6.6 & 2.21$\substack{0.56 \\ -0.49}$ \\
N2CO\_1\_114 & N2CO J100103+022140 & 150.2639 & 2.3616 & 6.3 & 0.56$\substack{0.16 \\ -0.14}$ & 6.5 & 2.07$\substack{0.55 \\ -0.48}$ \\
N2CO\_1\_118 & N2CO J100035+022827 & 150.1477 & 2.4740 & 6.1 & 0.57$\substack{0.17 \\ -0.14}$ & 6.4 & 2.04$\substack{0.53 \\ -0.46}$ \\
N2CO\_1\_128 & N2CO J100048+023016 & 150.2042 & 2.5033 & 6.1 & 0.57$\substack{0.16 \\ -0.14}$ & 6.2 & 1.71$\substack{0.50 \\ -0.45}$ \\
N2CO\_1\_138 & N2CO J095958+020604 & 149.9915 & 2.0998 & 5.2 & 0.49$\substack{0.18 \\ -0.15}$ & 5.9 & 1.98$\substack{0.59 \\ -0.50}$ \\
N2CO\_1\_143 & N2CO J095948+022752 & 149.9546 & 2.4641 & 5.7 & 0.54$\substack{0.16 \\ -0.15}$ & 5.9 & 1.94$\substack{0.59 \\ -0.50}$ \\
N2CO\_1\_156 & N2CO J100105+023239 & 150.2741 & 2.5447 & 4.7 & 0.46$\substack{0.18 \\ -0.14}$ & 5.6 & 1.89$\substack{0.57 \\ -0.50}$ \\
N2CO\_1\_166 & N2CO J100015+020531 & 150.0630 & 2.0921 & 6.6 & 0.61$\substack{0.17 \\ -0.15}$ & 5.4 & 1.72$\substack{0.56 \\ -0.45}$ \\
N2CO\_1\_167 & N2CO J100006+023306 & 150.0281 & 2.5519 & 4.8 & 0.48$\substack{0.19 \\ -0.15}$ & 5.4 & 1.90$\substack{0.60 \\ -0.51}$ \\
N2CO\_1\_170 & N2CO J100057+021347 & 150.2373 & 2.2284 & 5.0 & 0.46$\substack{0.18 \\ -0.14}$ & 5.3 & 1.71$\substack{0.54 \\ -0.46}$ \\
N2CO\_1\_204 & N2CO J100119+022617 & 150.3303 & 2.4385 & 4.7 & 0.50$\substack{0.20 \\ -0.16}$ & 4.8 & 1.81$\substack{0.67 \\ -0.55}$ \\
N2CO\_1\_208 & N2CO J100027+023344 & 150.1166 & 2.5632 & 5.2 & 0.54$\substack{0.21 \\ -0.17}$ & 4.8 & 1.67$\substack{0.62 \\ -0.52}$ \\
N2CO\_2\_29\tablefootmark{b} & N2CO J100001+022819 & 150.0083 & 2.4721 & 7.1 & 0.67$\substack{0.15 \\ -0.14}$ &  -- & --\\
N2CO\_2\_37 & N2CO J100114+020208 & 150.3123 & 2.0356 & 6.6 & 0.71$\substack{0.20 \\ -0.17}$ &  -- & $<$2.71\\
N2CO\_2\_44 & N2CO J095945+023438 & 149.9380 & 2.5773 & 6.3 & 0.79$\substack{0.22 \\ -0.19}$ &  -- & $<$2.92\\
N2CO\_2\_49 & N2CO J100100+020923 & 150.2523 & 2.1566 & 6.1 & 0.57$\substack{0.16 \\ -0.14}$ &  -- & $<$2.68\\
N2CO\_2\_61\tablefootmark{b} & N2CO J095952+020544 & 149.9671 & 2.0957 & 5.5 & 0.53$\substack{0.16 \\ -0.14}$ &  -- & --\\
N2CO\_2\_76 & N2CO J100042+020739 & 150.1759 & 2.1276 & 4.9 & 0.46$\substack{0.18 \\ -0.15}$ &  -- & $<$2.24\\
N2CO\_2\_77 & N2CO J095953+020734 & 149.9735 & 2.1261 & 4.9 & 0.45$\substack{0.18 \\ -0.14}$ &  -- & $<$3.17\\
N2CO\_2\_85 & N2CO J100059+022603 & 150.2489 & 2.4343 & 4.7 & 0.43$\substack{0.17 \\ -0.14}$ &  -- & $<$2.82\\
N2CO\_2\_88 & N2CO J100050+022953 & 150.2103 & 2.4983 & 4.6 & 0.41$\substack{0.17 \\ -0.13}$ &  -- & $<$2.51\\
N2CO\_2\_90 & N2CO J095944+022227 & 149.9346 & 2.3744 & 4.6 & 0.45$\substack{0.18 \\ -0.15}$ &  -- & $<$3.10\\
\hline
\end{tabular}
\tablefoot{
\tablefoottext{b}{These two 2\,mm sources are complex blends with a slightly-offset bright 1.2\,mm counterpart and potentially a fainter emission coming from the 2\,mm object (see Carvajal- Bohorquez et al. in prep.). For this reason, we cannot provide an upper limit on their 1.2\,mm flux.}}
\end{table*}

\end{appendix}

\end{document}